\documentclass[11pt]{article}
\usepackage[utf8]{inputenc}
\usepackage[margin=1.25in]{geometry}
\usepackage{amsmath}
\usepackage{amssymb}
\usepackage{amsthm}
\usepackage{bbm}

\usepackage{multicol}
\usepackage{multirow}
\usepackage{booktabs}
\usepackage{graphicx}
\usepackage{float}
\usepackage{bm}
\usepackage{multirow}
\usepackage{xargs}
\usepackage{listings}
\usepackage{xcolor}
\usepackage{tabularx}
\usepackage{algorithm}
\usepackage{algcompatible}
\usepackage{rotating}
\usepackage{comment}

\usepackage{bbm}
\usepackage{comment}
\usepackage{natbib}
\usepackage[colorlinks=true, allcolors=blue]{hyperref}

\theoremstyle{definition}

\newcommand{\rightarrowp}{\xrightarrow{p}}

\newcommand{\footremember}[2]{
    \footnote{#2}
    \newcounter{#1}
    \setcounter{#1}{\value{footnote}}
}
\newcommand{\footrecall}[1]{
    \footnotemark[\value{#1}]
} 

\providecommand{\keywords}[1]
{
  \small	
  \textbf{\textit{Keywords---}} #1
}

\title{Prognostic Covariate Adjustment for Logistic Regression in Randomized Controlled Trials
\author{Yunfan Li\footremember{Karuna}{Senior Principal Statistician, Karuna Therapeutics, Boston, MA, USA, yunfan.li@karunatx.com}, \and Arman Sabbaghi\footnotemark\footnote{Corresponding author} \footremember{Unlearn.AI}{Unlearn.AI, Inc., San Francisco, CA, USA}, \and Jonathan R. Walsh\footrecall{Unlearn.AI}, \and Charles K. Fisher\footrecall{Unlearn.AI}}
\date{\today}}

\begin{document}

\maketitle

\begin{abstract}
Randomized controlled trials (RCTs) with binary primary endpoints introduce novel challenges for inferring the causal effects of treatments. The most significant challenge is non-collapsibility, in which the conditional odds ratio estimand under covariate adjustment differs from the unconditional estimand in the logistic regression analysis of RCT data. This issue gives rise to apparent paradoxes, such as the variance of the estimator for the conditional odds ratio from a covariate-adjusted model being greater than the variance of the estimator from the unadjusted model. We address this challenge in the context of adjustment based on predictions of control outcomes from generative artificial intelligence (AI) algorithms, which are referred to as prognostic scores. We demonstrate that prognostic score adjustment in logistic regression increases the power of the Wald test for the conditional odds ratio under a fixed sample size, or alternatively reduces the necessary sample size to achieve a desired power, compared to the unadjusted analysis. We derive formulae for prospective calculations of the power gain and sample size reduction that can result from adjustment for the prognostic score. Furthermore, we utilize g-computation to expand the scope of prognostic score adjustment to inferences on the marginal risk difference, relative risk, and odds ratio estimands. We demonstrate the validity of our formulae via extensive simulation studies that encompass different types of logistic regression model specifications. Our simulation studies also indicate how prognostic score adjustment can reduce the variance of g-computation estimators for the marginal estimands while maintaining frequentist properties such as asymptotic unbiasedness and Type I error rate control. Our methodology can ultimately enable more definitive and conclusive analyses for RCTs with binary primary endpoints.
\end{abstract}


\section{Introduction}
\label{sec:introduction}

Randomized controlled trials (RCTs) serve as the gold standard for causal inferences on new medical treatments and interventions \citep{bhide_et_al_2018}. A fundamental consideration in the design and analysis of a RCT is the primary endpoint, which is used to measure the evidence for the primary objective of the RCT \citep[p.~7--8]{food_and_drug_administration_ich_1998}. Binary outcomes are commonly considered for primary endpoints \citep{rombach_et_al_2020}. A binary endpoint has only two possible values, typically denoted by $0$ and $1$ and referred to as ``non-event'' and ``event'', respectively. It can correspond to a coarsened version of a latent, continuous measure \citep{qu_et_al_1992, hu_et_al_2004}. Consequently, binary endpoints typically provide less information for treatment effect inferences compared to continuous endpoints \citep{schmitz_et_al_2012}. Key objectives in the design and analysis of RCTs with binary endpoints are to increase the precision of the treatment effect estimator and the power for the hypothesis test on the treatment effect, while preserving asymptotic unbiasedness of the treatment effect estimator and control of the Type I error rate for the hypothesis test.

Increasing the RCT sample size to achieve these objectives is a costly and time-consuming endeavor. A more practical approach is to perform covariate adjustment \citep{food_and_drug_administration_adjusting_2023}. A recent demonstration of the utility of covariate adjustment is the testing of COVID-19 treatments \citep{benkeser2021improving}. Covariate adjustment in the analysis of a binary endpoint is typically performed via logistic regression \citep{berkson_1944}, in which a logistic function is fit to the data to model the probability of an event as a function of the treatment indicator and the covariate(s). Beyond binary endpoints, covariate adjustment has been used to great effect in the analysis of continuous primary endpoints in RCTs. This is because covariates that are highly associated with the outcome generally yield treatment effect inferences with substantially greater precision and power compared to unadjusted analyses \citep{schuler2022increasing}. An exciting and modern approach to covariate adjustment is the use of generative artificial intelligence (AI) algorithms to construct digital twins for trial participants. These algorithms can be pre-trained on historical control data to summarize the high-dimensional covariate vectors for RCT participants into a one-dimensional prediction of their control outcomes. This prediction is referred to as a prognostic score \citep{hansen_2008}. \citet{schuler2022increasing} developed a statistical methodology for prognostic covariate adjustment of continuous endpoints, referred to as PROCOVA$^{\mathrm{TM}}$. This method is qualified by the European Medicines Agency (EMA) as ``an acceptable statistical approach for primary analysis'' of Phase 2 and 3 RCTs with continuous endpoints \citep{ema_procova_2022}.

Covariate adjustment for binary endpoints introduces novel challenges for causal inferences in RCTs. A significant challenge that has been highlighted by the \citet[FDA,][]{food_and_drug_administration_adjusting_2023} is non-collapsibility, or the discrepancy between the conditional treatment effect that incorporates the covariate adjustment and the unconditional, or marginal, treatment effect for a population of RCT participants \citep{greenland2021noncollapsibility, daniel2021making}. In particular, non-collapsibility complicates comparisons of precisions for the treatment effect estimators from unadjusted and adjusted logistic regression models. For example, a consequence of non-collapsibility is the seemingly paradoxical result that covariate adjustment leads to variance inflation of the treatment effect estimator compared to the unadjusted analysis \citep{robinson1991some, daniel2021making}. This consequence has unfortunately created a great deal of confusion and misunderstanding about the utility of covariate adjustment for the analysis of binary endpoints in RCTs.

We address this challenge in the context of adjustment for an AI-generated prognostic score in the analysis of binary endpoints. Our methodology is referred to as prognostic covariate adjustment for logistic regression, and is abbreviated as PROCOVA-LR. We derive new formulae to establish that PROCOVA-LR can increase the power of the Wald test for the conditional odds ratio estimand compared to the unadjusted logistic regression analysis for a fixed sample size, and that PROCOVA-LR can reduce the necessary sample size to achieve a desired power compared to the unadjusted analysis. Our formulae can be prospectively calculated prior to the RCT, and is a simple function of the average of the participants' predicted probabilities of an event under control and the variance of their predicted probabilities, with the prognostic scores used to calculate the predicted probabilities. We further expand the scope of prognostic score adjustment to g-computation inferences on the marginal risk difference, relative risk, and odds ratio estimands \citep{freedman2008randomization, steingrimsson_2017}. In particular, we demonstrate that PROCOVA-LR can lead to estimators for these estimands with reduced variance compared to those from the unadjusted logistic regression analysis. 

We proceed in Section \ref{sec:background} to provide the notations, assumptions, and the framework for PROCOVA-LR. Our prospective formulae for power gains and sample size reductions for the Wald test on the conditional estimand, calculations of the variances of the g-computation estimators for the marginal estimands, and derivations of the Wald test statistics for the tests on the marginal estimands are in Section \ref{sec:procova_lr}. These theoretical results are validated via extensive simulation studies, spanning both well-specified and misspecified PROCOVA-LR models relative to the true data generation mechanism, in Section \ref{sec:sim}. A key result from these simulation studies is that PROCOVA-LR can yield more powerful tests on the estimands and more efficient inferences on the marginal estimands compared to the unadjusted logistic regression analysis, while controlling the asymptotic bias of the estimators and Type I error rates of the tests. Our concluding remarks are in Section \ref{sec:disc}. Ultimately, PROCOVA-LR addresses the challenge of non-collapsibility in the analysis of binary endpoints for RCTs, providing meaningful, powerful, and interpretable inferences for multiple treatment effects of interest by leveraging the power of modern AI for covariate adjustment.

\section{Background}
\label{sec:background}

\subsection{Notations and Assumptions}
\label{sec:notation_assumptions}

We adopt the Neyman-Rubin Causal Model \citep{neyman1923causal, rubin_1974, holland1986statistics} to define the experimental units, treatment indicators, covariates, and potential outcomes for an RCT. These elements are essential to describe the traditional, conditional odds ratio estimand, as well as the marginal estimands for binary endpoints. The marginal estimands are the risk difference (RD), relative risk (RR), and odds ratio (OR). After defining these elements of the Neyman-Rubin Causal Model, in the remainder of this section we review logistic regression, non-collapsibility, bias factors and asymptotic relative efficiencies (AREs) of logistic regression coefficient estimators, and g-computation for inferring the marginal estimands.

An experimental unit corresponds to a single participant in an RCT at a specified point in time \citep[p.~4]{imbens_rubin_2015}. We use the terms ``experimental unit'' and ``participant'' interchangeably. For each participant $i = 1, \ldots, N$, we let $w_i \in \{0, 1\}$ indicate their treatment assignment and $x_i \in \mathbb{R}^L$ denote their covariate vector. We indicate assignment of the active treatment to participant $i$ by $w_i = 1$, and assignment of the control (e.g., a placebo) by $w_i = 0$. Each participant can only be assigned to, and receive, at most one treatment level, i.e., they cannot be assigned to both the active treatment and control. A participant's covariate vector contains their characteristics that are observed either prior to treatment assignment, or after treatment assignment and are known to be unaffected by treatment \citep[p.~15--16]{imbens_rubin_2015}. The possible values of the binary endpoint are denoted by $0$ and $1$, with $1$ indicating an event. We invoke the Stable Unit-Treatment Value Assumption \citep[SUTVA,][p.~9--13]{imbens_rubin_2015}, in which each combination of participant and treatment level corresponds to a well-defined binary outcome, and the outcome for a participant does not depend on treatments assigned to others. Under SUTVA, the potential outcome for participant $i$ under treatment $w$ is unambiguously defined by $Y_i(w)$. 

Marginal causal estimands are defined via a comparison of the $Y_i(1)$ versus the $Y_i(0)$ for a population of participants. An example is the risk difference for the RCT, $\Delta_{\mathrm{RD}} = \bar{Y}(1) - \bar{Y}(0) = N^{-1} \left \{ \sum_{i=1}^N Y_i(1) - \sum_{i=1}^N Y_i(0) \right \}$. Two other marginal estimands for the RCT are the relative risk $\Delta_{\mathrm{RR}} = \bar{Y}(1)/\bar{Y}(0)$ and the odds ratio $\Delta_{\mathrm{OR}} = \left [ \bar{Y}(1)/ \left \{ 1-\bar{Y}(1) \right \} \right ] / \left [ \bar{Y}(0)/ \left \{ 1 - \bar{Y}(0) \right \} \right ]$. For the latter two estimands we assume $\bar{Y}(w) \neq 0, 1$ for $w \in \{0, 1\}$.

Causal inference under the Neyman-Rubin Causal Model is a missing data problem, with at most one potential outcome observed for any participant \citep[p.~38]{rubin_1978}. The observed outcomes are functions of the treatment indicators and potential outcomes via $y_i = w_iY_i(1) + \left ( 1 - w_i \right )Y_i(0)$.  The treatment assignment mechanism, i.e., the probability mass function $p ( w_1, \ldots, w_N \mid Y_1(0), \ldots, Y_N(0)$, $Y_1(1), \ldots, Y_N(1),$ $x_1, \ldots, x_N)$, corresponds to a missing data mechanism \citep[p.~43]{imbens_rubin_2015}. Three important regularity conditions for a treatment assignment mechanism are that it is unconfounded (i.e., there are no lurking confounders associated with both treatment assignment and the potential outcomes conditional on the covariates), probabilistic, and individualistic (i.e., a participant's treatment assignment does not depend on the covariates or potential outcomes of others) \citep[p.~37--39]{imbens_rubin_2015}. Violations of these regularity conditions would complicate the design and analysis of an RCT. The completely randomized design is an assignment mechanism for RCTs that satisfies these regularity conditions.

\subsection{Logistic Regression}
\label{sec:logistic_regression}

Logistic regression is an established methodology for modeling the probability of an event for a binary endpoint as a function of predictor variables \citep{mccullagh_nelder_1989, faraway_2016}. The (unknown) probabilities $\mathrm{Pr} \left \{ Y_i(1) = 1 \mid x_i \right \}$ and $\mathrm{Pr} \left \{ Y_i(0) = 1 \mid x_i \right \}$ are modeled based on the observed outcomes $y_i$ and the application of the standard logistic function to the dot product of a vector of predictors $v_i \in \mathbb{R}^K$ (defined based on the $w_i$ and $x_i$) and unknown regression coefficients $\beta = \left ( \beta_0, \ldots, \beta_{K-1} \right )^{\mathsf{T}} \in \mathbb{R}^K$. Traditional inferences on the conditional odds ratio estimand via logistic regression involve inferences for the entry in $\beta$ corresponding to $w_i$. Under the Neyman-Rubin Causal Model, inferences on $\Delta_{\mathrm{RD}}, \Delta_{\mathrm{RR}}$, and $\Delta_{\mathrm{OR}}$ can also be performed by combining logistic regression with either multiple imputation of missing potential outcomes \citep[p.~1799--1800]{pattanayak2012cmh, gutman_2013} or g-computation \citep{freedman2008randomization}. 

The logistic regression model is fitted to the observed outcomes and predictors via maximum likelihood estimation. The general model specification is $\mathrm{Pr} \left ( y_i = 1 \mid v_i \right )  = \mathrm{exp} \left (v_i^{\mathsf{T}}\beta \right )/ \left \{ 1 + \mathrm{exp} \left (v_i^{\mathsf{T}}\beta \right ) \right \}$. All potential outcomes are assumed to be mutually independent conditional on the predictors. The corresponding general likelihood function is $L \left ( \beta \right ) = \displaystyle \prod_{i=1}^N \left [ \mathrm{exp} \left ( w_iv_i^{\mathsf{T}}\beta \right ) \left \{ 1 + \mathrm{exp} \left ( v_i^{\mathsf{T}}\beta \right ) \right \}^{-1} \right ]$. We assume that there is no complete or quasi-complete separation, that the endpoint values are not sparse, and that there is no perfect collinearity in the matrix of predictor vectors $V = \begin{pmatrix} v_1^{\mathsf{T}} \\ \vdots \\ v_N^{\mathsf{T}} \end{pmatrix}$. Under these assumptions, maximum likelihood-based inferences can be performed for logistic regression  \citep[p.~117, 120--122]{mccullagh_nelder_1989}. 

The interpretation of the parameters in $\beta$ depend on the predictors in $v_i$. For example, consider $v_i = \left ( 1, w_i \right )^{\mathsf{T}}$, which corresponds to the unadjusted logistic regression model. In this case we denote the entries in $\beta$ by $\beta_0^*$ and $\beta_1^*$, and interpret $\mathrm{exp} \left ( \beta_0^* \right )$ as the odds of an event under control and $\mathrm{exp} \left ( \beta_1^* \right )$ as the multiplicative change in the odds of an event under treatment compared to control. The latter coefficient is a type of marginal estimand, as the model does not condition on any covariates or predictors besides $w_i$. The unadjusted logistic regression model is
\begin{equation}
\label{eq:unadj}
\mathrm{Pr} \left ( y_i = 1 \mid v_i \right ) = \frac{\mathrm{exp} \left ( \beta_0^* + \beta_1^*w_i \right )}{1 + \mathrm{exp} \left ( \beta_0^* + \beta_1^*w_i \right )}.
\end{equation}
These interpretations differ from the case in which additional predictors are included in $v_i$. To illustrate, now consider the case in which $v_i = \left ( 1, w_i, x_i \right )^{\mathsf{T}}$ includes the covariate $x_i \in \mathbb{R}$ in addition to the treatment indicator. This is an adjusted logistic regression model. We denote the entries in $\beta$ by $\beta_0, \beta_1$, and $\beta_2$ in this case, and the model specification is
\begin{equation}
\label{eq:adj}
\mathrm{Pr} \left ( y_i = 1 \mid v_i \right ) = \frac{\mathrm{exp} \left ( \beta_0 + \beta_1 w_i + \beta_2 x_i \right )}{1 + \mathrm{exp} \left ( \beta_0 + \beta_1 w_i + \beta_2 x_i \right )}.
\end{equation}
Now $\mathrm{exp} \left (\beta_0 \right )$ is the odds of an event under control when $x_i = 0$, and $\mathrm{exp} \left ( \beta_1 \right )$ is the multiplicative change in the odds of an event under treatment compared to control. The estimand $\mathrm{exp} \left ( \beta_1 \right )$ is defined conditional on $x_i$, and differs from $\mathrm{exp} \left ( \beta_1^* \right )$ because the latter estimand is effectively calculated by averaging over the distribution of the covariate $x_i$ \citep[p.~528--529]{daniel2021making}. Other logistic regression model specifications can be considered, such as those that include multiple covariates, transformations of covariates, and/or interactions between the treatment indicator and covariates. Guidance documents published by regulatory agencies recommend that the number of predictors in adjusted analyses be kept to an appropriate minimum \citep{ema_adjusting_2015, food_and_drug_administration_adjusting_2023}. 

The $\beta_1^*$ and $\beta_1$ parameters in equations (\ref{eq:unadj}) and (\ref{eq:adj}) have traditionally been referred to as ``treatment effects'' for the unadjusted and adjusted models, respectively. Under the Neyman-Rubin Causal Model, these are not valid finite-population treatment effects as they do not involve the potential outcomes for the participants \citep[p.~18]{imbens_rubin_2015}. Furthermore, these parameters have different interpretations and magnitudes, because $\beta_1^*$ is defined without consideration of the covariate whereas $\beta_1$ is defined conditional on $x_i$. In contrast, treatment effects for binary endpoints can be unambiguously defined in a manner that is agnostic to the logistic regression model specification by considering the marginal estimands $\Delta_{\mathrm{RD}}, \Delta_{\mathrm{RR}}$, and $\Delta_{\mathrm{OR}}$. The \citet[p.~6]{food_and_drug_administration_adjusting_2023} guidance on covariate adjustment notes that, for a statistical analysis involving a nonlinear model (e.g., logistic regression), sponsors should discuss their plans to analyze a conditional estimand such as $\beta_1$ in the primary analysis. Furthermore, this document notes that sponsors can perform covariate-adjusted estimation and inference for marginal estimands, such as $\Delta_{\mathrm{RD}}, \Delta_{\mathrm{RR}}$, and $\Delta_{\mathrm{OR}}$. This serves to indicate that both types of estimands can be considered by regulators.

\subsection{Non-Collapsibility in Logistic Regression}
\label{sec:noncollapsibility_logistic_regression}

The phenomenon in which the definition of an estimand depends on the covariates that are included in the statistical analysis is non-collapsibility \citep{agresti_2002, daniel2021making}. For a binary endpoint, the odds ratio estimands are non-collapsible, whereas the risk difference and relative risk estimands are collapsible \citep[p.~5--6]{food_and_drug_administration_adjusting_2023}. Historically, the consequences of this phenomenon created a great deal of misunderstanding regarding the utility of covariate adjustment in logistic regression.

\citet[p.~7]{colnet_2023} characterize non-collapsibility as the situation in which the marginal estimand for a population cannot be expressed as a weighted combination of conditional estimands that are defined according to specified subpopulations. Alternatively, an estimand is collapsible if it can be expressed as such a weighted combination. According to \citet[p.~4,6]{colnet_2023}, collapsibility of an estimand is needed to generalize conditional estimands to a larger population. Also, collapsible marginal estimands typically do not require significant modeling assumptions for their definitions, whereas non-collapsible conditional estimands require a well-specified model in order to be well-defined. Besides non-collapsibility, estimands that are not logic-respecting according to the definition provided by \citet[p.~8]{colnet_2023} further complicate causal inferences. In particular, the odds ratio is neither collapsible nor logic-respecting, and \citet[p.~8]{colnet_2023} describe how the paradoxes associated with the odds ratio estimand are more attributable to the fact that it is not logic-respecting.

As an illustration of one significant complication resulting from non-collapsibility in logistic regression, consider the unadjusted model (\ref{eq:unadj}) and the adjusted model (\ref{eq:adj}). The latter model is generally of more importance in practice. However, a consequence of non-collapsibility is that the precision of the odds ratio estimator from model (\ref{eq:adj}) is less than that from model (\ref{eq:unadj}). Alternatively, the variance for the maximum likelihood estimator (MLE) $\hat{\beta}_1$ from model (\ref{eq:adj}) could be greater than that for the MLE $\widehat{\beta_1^*}$ from model (\ref{eq:unadj}) \citep{robinson1991some}. This inequality in the estimators' precisions is difficult to reconcile with the fact that the Wald test for $H_0: \beta_1 = 0$ under logistic regression with covariate adjustment could have more power than that for $H_0: \beta_1^* = 0$ without covariate adjustment \citep{robinson1991some}. It also contradicts one's intuition from linear regression (which involves a collapsible estimand), in which covariate adjustment leads to an increase in both the precision for the coefficient estimator and the power for testing the coefficient. The seemingly paradoxical relationship between a coefficient estimator's precision and the power of the Wald test previously deterred researchers from adopting covariate adjustment in logistic regression.

This paradox is explained by the fact that the two estimands from models (\ref{eq:unadj}) and (\ref{eq:adj}) differ in both nature and magnitude, with the magnitude of the conditional odds ratio generally being greater than that for the unconditional odds ratio \citep{robinson1991some}. Besides the different magnitudes for these two estimands, concerns about the power discrepancy can also be resolved by realizing that when there is no treatment effect, the population (marginal) and the subgroup (conditional) odds ratios both equal $1$, and the odds ratio is strictly collapsible \citep{didelez2022logic}.

\subsection{Bias Factors and Asymptotic Relative Efficiencies of Logistic Regression Coefficient Estimators}
\label{sec:bias_factor_ARE}

\citet{neuhaus1998estimation} compared the coefficient estimators for models (\ref{eq:unadj}) and (\ref{eq:adj}) by calculating the (asymptotic) ``bias factor'' and asymptotic relative efficiency (ARE) of $\widehat{\beta_1^*}$ relative to  $\hat{\beta}_1$. These quantify the consequences of omitting a covariate that is associated with the outcome, i.e., of inferring the odds ratio estimand based on the unadjusted model when the adjusted model is more appropriate. This comparison is formulated in terms of the limiting case $\beta_1 \rightarrow 0$, which is relevant for the hypothesis test $H_0: \beta_1 = 0$. The work of \citet{neuhaus1998estimation} establishes that both the bias factor and ARE of $\widehat{\beta_1^*}$ are functions of $\beta_0$, $\beta_2$, and $x_i$, but not of $\beta_1$. This helps to explain the paradoxes of non-collapsibility.

To formally define the bias factor and ARE for $\widehat{\beta_1^*}$ versus $\hat{\beta}_1$, we first abuse notation and define the function $\beta_1^*: \mathbb{R}^3 \rightarrow \mathbb{R}$ in terms of the adjusted model (\ref{eq:adj}) and its parameters $\beta_0, \beta_1$, and $\beta_2$ by integrating over the distribution of the covariate $x_i$ as in \citep[p.~1126]{neuhaus1998estimation}. The bias factor is defined as
\begin{equation}
\label{eq:bias_factor_definition}
\lim_{\beta_1 \rightarrow 0} \frac{\partial}{\partial \beta_1} \left \{ \beta_1^* \left ( \beta_0, \beta_1, \beta_2 \right ) \right \},
\end{equation}
and captures the size of the treatment indicator coefficient under the misspecified, unadjusted model compared to the coefficient under the correctly specified, adjusted model near the value of $0$. As the bias factor considers only the case of $\beta_1 \rightarrow 0$, the linear term of the Taylor expansion of $\beta_1^*\left ( \beta_0, \beta_1, \beta_2 \right )$ about $\beta_1 = 0$ is sufficient to facilitate the calculation of equation (\ref{eq:bias_factor_definition}). \citet[p.~1126--1127]{neuhaus1998estimation} calculated the bias factor as
\begin{align}
\label{eq:bias_factor}
\lim_{\beta_1 \rightarrow 0} \frac{\partial}{\partial \beta_1} \left \{ \beta_1^* \left ( \beta_0, \beta_1, \beta_2 \right ) \right \} = 1 - \frac{\mathrm{Var}(\mu_{0,i})}{E(\mu_{0,i})\left \{ 1-E(\mu_{0,i}) \right \} }, 
\end{align}
where $\mu_{0,i} = \mathrm{exp} \left ( \beta_0 + \beta_2 x_i \right ) / \left \{ 1 + \mathrm{exp} \left ( \beta_0 + \beta_2 x_i \right ) \right \}$ denotes the predictive probability of an event for participant $i$ under control. If the variance of the $x_i$ is $0$, or if $\beta_2 = 0$, then the bias factor is zero and $\beta_1 = \beta_1^*$ as expected. The bias factor increases as a function of the variance in $x_i$ and/or the value of $\beta_2$.

The ARE of $\widehat{\beta_1^*}$ versus $\widehat{\beta}_1$ evaluated at $0$ is defined by \citet[p.~1125]{neuhaus1998estimation} as
\begin{equation}
\label{eq:ARE_definition}
\mathrm{ARE}(\widehat{\beta_1^*} , \hat{\beta}_1) = \biggl[\lim_{\beta_1 \rightarrow 0} \biggl\{\frac{\partial \beta_1^*}{\partial \beta_1}\biggr\} \biggl\{\frac{\partial \beta_1}{\partial \beta_1}\biggr\}^{-1} \biggr]^2 \biggl[\lim_{\beta_1 \rightarrow 0} \frac{\mathrm{Var}(\hat{\beta}_1)}{\mathrm{Var}(\widehat{\beta_1^*})}  \biggr].
\end{equation}
Under $H_0: \beta_1 = 0$, and by virtue of the independence of treatment assignment and the covariate in an RCT, expressions for $\mathrm{Var}(\hat{\beta}_1)$ and $\mathrm{Var}(\widehat{\beta_1^*})$ are obtained that involve only the expectation and variance of $\mu_{0,i}$. \citet[p.~1127]{neuhaus1998estimation} then derived the ARE as
\begin{equation}
\label{eq:ARE}
\mathrm{ARE}(\widehat{\beta_1^*} \, \mathrm{to} \, \hat{\beta}_1 \, \mathrm{at} \, \beta_1=0) = 1-\frac{\mathrm{Var}(\mu_{0,i})}{E(\mu_{0,i}) \left \{ 1 - E(\mu_{0,i}) \right \} },
\end{equation}
which is equivalent to equation (\ref{eq:bias_factor}). The ARE is always less than $1$ when $\mathrm{Var} \left ( \mu_{0,i} \right ) > 0$, so that the estimator for the conditional odds ratio estimand from the unadjusted model has a smaller variance than the estimator from the adjusted model. Although this appears to be a paradox for covariate adjustment in logistic regression, we demonstrate in Section \ref{sec:procova_lr} how equations (\ref{eq:bias_factor}) and (\ref{eq:ARE}) establish that covariate adjustment can increase the power of the Wald test for the conditional odds ratio estimand in logistic regression.

\subsection{G-Computation for Inferring Marginal Estimands on Binary Endpoints}
\label{sec:g-computation}

We utilize g-computation to infer $\Delta_{\mathrm{RD}}, \Delta_{\mathrm{RR}}$, and $\Delta_{\mathrm{OR}}$ under the Neyman-Rubin Causal Model. This approach was developed by \citet{freedman2008randomization}, and has been recognized as a valid statistical method of covariate adjustment to infer marginal estimands in the case of binary endpoints \citep[p.~7]{ge_2011, snowden2011implementation, food_and_drug_administration_adjusting_2023}. It meets the criteria of the \citet{food_and_drug_administration_adjusting_2023} for inferring marginal estimands based on covariate adjustment, as it provides valid causal inferences under the same type of minimal statistical assumptions that would be involved for unadjusted analyses.

G-computation is effectively a ``plug-in'' estimator that utilizes the MLEs of the logistic regression coefficients to replace all observed and missing potential outcomes in the RCT by their predicted probabilities \citep[p.~1799]{gutman_2013}. It yields consistent estimators for the interpretable causal estimands in the case of non-collapsibility under the logistic regression model, even in the case of model misspecification \citep[p.~3--5]{freedman2008randomization}. Furthermore, its statistical efficiency is supported by asymptotic theory \citep{rosenblum_2016}. It can also be performed for stratified designs by extending the simple method-of-moments estimators and Cochran-Mantel-Haenszel-Tarone estimator described by \citet{graf_schumacher_2008}, \citet{stampf_et_al_2010}, and \citet[p.~4--5]{pattanayak2012cmh} using a fitted logistic regression model. Another advantage of g-computation is that it is model agnostic, in that it targets estimands that are well-defined in terms of potential outcomes without reference to any specified model. This enables researchers to infer estimands of scientific interest and frees them from selecting estimands based on mathematical convenience or modeling conventions. For example, if the risk difference is pertinent, then $\Delta_{\mathrm{RD}}$ rather than $\Delta_{\mathrm{OR}}$ can be the target estimand, and any model can be utilized to infer it via g-computation.

As an illustration of g-computation, consider the application of model (\ref{eq:adj}) to infer the marginal risk difference, relative risk, and odds ratio estimands. Let $\hat{\beta} = \left ( \hat{\beta}_0, \hat{\beta}_1, \hat{\beta}_2 \right )^{\mathsf{T}}$ denote the MLEs for the logistic regression coefficients. For each participant we estimate their probabilities of an event under treatment and control by $p_i \left ( 1 \right ) = \mathrm{exp} \left ( \hat{\beta}_0 + \hat{\beta}_1 + \hat{\beta}_2 x_i \right )/  $ $\left \{ 1 + \mathrm{exp} \left ( \hat{\beta}_0 + \hat{\beta}_1 + \hat{\beta}_2 x_i \right ) \right \}$ and $p_i \left ( 0 \right ) = \mathrm{exp} \left ( \hat{\beta}_0 + \hat{\beta}_2 x_i \right ) / \left \{ 1 + \mathrm{exp} \left ( \hat{\beta}_0 + \hat{\beta}_2 x_i \right ) \right \}$, respectively. We then use these estimators to calculate the averages $\bar{p}(1) = \sum_{i=1}^N p_i(1)/N$ and $\bar{p}(0) = \sum_{i=1}^N p_i(0)/N$. Finally, the point estimators of $\Delta_{\mathrm{RD}}$, $\Delta_{\mathrm{RR}}$, and $\Delta_{\mathrm{OR}}$ are obtained by replacing $\bar{Y}(1)$ by $\bar{p}(1)$ and $\bar{Y}(0)$ by $\bar{p}(0)$ in the definitions of the original estimands, as in plug-in estimation. More formally, $\widehat{\Delta_{\mathrm{RD}}} = \bar{p}(1) - \bar{p}(0)$, $\widehat{\Delta_{\mathrm{RR}}} = \bar{p}(1)/\bar{p}(0)$, and $\widehat{\Delta_{\mathrm{OR}}} = \left [ \bar{p}(1)/ \left \{ 1 - \bar{p}(1) \right \} \right ] / \left [ \bar{p}(0)/ \left \{ 1 - \bar{p}(0) \right \} \right ]$.

Alternative methods exist to g-computation. One set of alternatives is based on multiple imputation of missing potential outcomes, with specific methods provided by \citet[p.~5--7]{pattanayak2012cmh} and \citet[p.~1799-1800]{gutman_2013}. Another alternative is described by \citet[p.~4]{freedman2008randomization} as an ``intention-to-treat'' estimator, which is a misnomer because it does not correspond to the definition of the intention-to-treat estimator in RCTs as described in \citep{food_and_drug_administration_ich_1998, food_and_drug_administration_ich_2021}. This approach is similar to simple method-of-moments estimation, e.g., the methods of \citet{graf_schumacher_2008} and \citet{stampf_et_al_2010} for stratified designs. It also results in consistent estimators \citep{robins1986new, freedman2008randomization, ye2023robust}.

\section{Prognostic Covariate Adjustment in Logistic Regression}
\label{sec:procova_lr}

\subsection{Adjustments for Prognostic Scores}
\label{sec:single_covariate}

Our PROCOVA-LR methodology conducts inferences for the conditional odds ratio estimand and the marginal risk difference, relative risk, and odds ratio estimands by adjusting for a single covariate in logistic regression. The single predictor is the prognostic score, and it is defined for each participant as the expectation of their respective digital twin distribution. The prognostic score effectively serves as a one-dimensional summary of the (potentially high-dimensional) baseline covariates, and can be highly associated with the probability of an event under control. It satisfies regulatory guidances on covariate adjustment that recommend a small number of covariates for adjustment \citep{ema_adjusting_2015, food_and_drug_administration_adjusting_2023}. In addition, fitting a logistic regression model that adjusts solely for the prognostic score instead of the high-dimensional covariate vector liberates degrees of freedom in the model.

The digital twin distribution for participant $i = 1, \ldots, N$ at a specified time-point is the Bernoulli$(m_{i})$ distribution for their potential outcome under control, with the probability $m_{i}$ of an event being a function of their covariate vector $x_i \in \mathbb{R}^L$. The prognostic score for participant $i$ is defined as $m_{i}$. This function is determined in practice by training the AI algorithm on an independent set of historical control data, separate from the RCT data. It can be calculated prospectively for an RCT, prior to any treatment assignments, in a similar manner as the prognostic covariate adjustment methodology of \citet{schuler2022increasing} for continuous endpoints. The functional form of $m_i$ can be implemented by any mathematical or computational means. AI algorithms are particularly powerful in this context because they can effectively capture associations between baseline predictors and the probability of an event under control. Furthermore, the rapid accumulation of historical control data and recent advances in AI further increase the promise and potential of prognostic scores for improving the quality of inferences via adjustment. The use of historical control data for modeling and validating the prognostic score helps to eliminate additional model selection steps in logistic regression that would complicate the analysis of an RCT.

The sole inputs for the AI algorithm that specifies the digital twin distribution are baseline covariates, and so the prognostic score itself is a covariate that can be incorporated as a predictor in logistic regression. The model for PROCOVA-LR is
\begin{equation}
\label{eq:procova_lr_model}
\mathrm{Pr} \left (y_i = 1 \mid w_i, m_i \right ) = \frac{\mathrm{exp} \left ( \beta_0 + \beta_1 w_i + \beta_2 m_i \right )}{1 + \mathrm{exp} \left ( \beta_0 + \beta_1 w_i + \beta_2 m_i \right )}.
\end{equation}
Alternatively, one could use the logit transformation of the $m_i$ as the predictor in the PROCOVA-LR model. In either case, the inclusion of the prognostic score predictor in model (\ref{eq:procova_lr_model}) can yield three potential advantages over the unadjusted model (\ref{eq:unadj}). First is reducing the necessary sample size such that the power of the test for $H_0: \beta_1 = 0$ is the same as the power of the test for $H_0: \beta_1^* = 0$ (with the latter power level assumed to be pre-specified, e.g., at $0.8$). Second is boosting the power of the test for $H_0: \beta_1 = 0$ compared to that of the test for $H_0: \beta_1^* = 0$ for a fixed sample size. We provide formulae for prospective estimation of sample size reduction and power boost in Section \ref{sec:sample_size}. Third is improving the precision and power of g-computation based inferences for $\Delta_{\mathrm{RD}}, \Delta_{\mathrm{RR}}$, and $\Delta_{\mathrm{OR}}$. We elaborate on the variance calculations and derivations of the Wald test statistics for the g-computation based inferences on the marginal estimands in Section \ref{sec:estimand}. These properties of PROCOVA-LR are demonstrated via simulation in Section \ref{sec:sim}.

\subsection{Sample Size Reduction and Power Gain for Testing the Conditional Odds Ratio Estimand}
\label{sec:sample_size}

Sample size reductions and power gains for testing the conditional odds ratio estimand under PROCOVA-LR can be prospectively estimated by combining two sets of expressions. The first set consists of the Wald test statistics $W_{\mathrm{UN}}$ and $W_{\text{P-LR}}$ for the hypotheses $H_0: \beta_1^* = 0$ and $H_0: \beta_1 = 0$ of the treatment indicator coefficients from models (\ref{eq:unadj}) and (\ref{eq:procova_lr_model}), respectively. The second set consists of the formulae for the bias factor and ARE of $\widehat{\beta_1^*}$ versus $\hat{\beta}_1$ from equations (\ref{eq:bias_factor}) and (\ref{eq:ARE}), respectively. Although \citet{neuhaus1998estimation} derived the latter two equations, he did not derive sample size reductions or power gains that can result from covariate adjustment in logistic regression. Furthermore, \citet{neuhaus1998estimation} only considered a single covariate, whereas our consideration of prognostic scores from digital twins address this drawback as multiple covariates are encoded into the prognostic score. A contribution of our work is that it can better inform the design of a RCT for performing hypothesis tests on the odds ratio estimand via logistic regression with covariate adjustment.

We proceed to express the ratio of $W_{\mathrm{UN}}$ and $W_{\text{P-LR}}$ in terms of the bias factor and ARE for $\widehat{\beta_1^*}$ versus $\hat{\beta}_1$ to derive our formulae for sample size reductions and power gains. The Wald statistic for testing $H_0: \theta = 0$ for an unknown parameter $\theta$ is defined in general as $W = N^{1/2}\hat{\theta}_N \hat{V}_N^{-1/2}$, where $N$ indicates the sample size, $\hat{\theta}_N$ is a point estimator of $\theta$ such that $\hat{\theta}_N \rightarrowp \theta$ as $N \rightarrow \infty$, and $\hat{V}_N/N$ is a consistent estimator of the asymptotic variance of $\hat{\theta}_N$ \citep[p.~81--83]{barndorff-nielsen_1994}. The term $\hat{V}_N^{-1}$ is interpreted as the average amount of information provided by each observation \citep[p.~82]{barndorff-nielsen_1994}.  Our consideration of $W_{\mathrm{UN}} / W_{\text{P-LR}}$ is motivated by the recognition that the bias factor in equation (\ref{eq:bias_factor}) approximates $\widehat{\beta_1^*}/\hat{\beta}_1$ \citep{neuhaus1993geometric}, and the ARE in equation (\ref{eq:ARE}) approximates the ratio of the variances of $\widehat{\beta_1^*}$ and $\hat{\beta}_1$ \citep{neuhaus1998estimation}. Hence, for fixed $N$,
\begin{equation}
\label{eq:efficiency_factor}
\frac{W_{\mathrm{UN}}}{W_{\text{P-LR}}} \approx \frac{\widehat{\beta_1^*}/\hat{\beta}_1}{\sqrt{\mathrm{Var} \left ( \widehat{\beta_1^*} \right ) / \mathrm{Var} \left ( \hat{\beta}_1 \right )}} \approx \sqrt{1 - \frac{\mathrm{Var} \left ( \mu_{0,i} \right )}{E \left ( \mu_{0,i} \right ) \left \{ 1 - E \left ( \mu_{0,i} \right ) \right \}}},
\end{equation}
where the $\mu_{0,i}$ are defined as in Section \ref{sec:bias_factor_ARE} but with $x_i$ replaced by $m_i$. The expectation and variance in this equation are calculated for the entire population of $\mu_{0,i}$ values. We refer to the right-hand side of equation (\ref{eq:efficiency_factor}) as the \emph{efficiency factor}, and denote it by $f_{\mathrm{EFF}}$. In general, a smaller value of $f_{\mathrm{EFF}}$ is better for PROCOVA-LR, as it indicates greater sample size reduction or power gain under PROCOVA-LR compared to the unadjusted analysis. This factor decreases as $\mathrm{Var} \left ( \mu_{0,i} \right )$ increases for fixed $E \left ( \mu_{0,i} \right )$. Under this situation, adjustment for the prognostic score will have a larger effect on the Wald test statistic, and hence the sample size reduction and power gain of PROCOVA-LR compared to the unadjusted model.

Our prospective (total) sample size reduction formula for powering a study with respect to PROCOVA-LR is derived by solving for $N_{\text{P-LR}}$ when $N_{\mathrm{UN}}$ is fixed in equation (\ref{eq:efficiency_factor}). To demonstrate this, we first recognize that Wald test statistics can yield approximations for power calculations. This is because the distributions of $W_{\mathrm{UN}}$ and $W_{\text{P-LR}}$ can be approximated by standard Normal distributions under their corresponding null hypotheses, and the powers of the Wald tests for the treatment indicator coefficients in models (\ref{eq:unadj}) and (\ref{eq:procova_lr_model}) can be approximated by
\begin{equation}
\label{eq:power_unadj}
\zeta_{\mathrm{UN}} = \Phi \biggl( \Phi^{-1} \biggl( \frac{\alpha}{2}\biggl)+W_{\mathrm{UN}}  \biggl) + \Phi \biggl( \Phi^{-1} \biggl( \frac{\alpha}{2}\biggl)-W_{\mathrm{UN}}  \biggl)
\end{equation}
and
\begin{equation}
\label{eq:power_adj}
\zeta_{\text{P-LR}} = \Phi \biggl( \Phi^{-1} \biggl( \frac{\alpha}{2}\biggl)+W_{\text{P-LR}}  \biggl) + \Phi \biggl( \Phi^{-1} \biggl( \frac{\alpha}{2}\biggl)-W_{\text{P-LR}} \biggl),
\end{equation}
respectively, where $\alpha$ is the Type I error rate (and usually taken as $0.05$). Thus, suppose $N_{\mathrm{UN}}$ is identified such that model (\ref{eq:unadj}) has power $0.8$ for testing $H_0: \beta_1^* = 0$. Using the previous points, we identify $N_{\text{P-LR}}$ such that the PROCOVA-LR model (\ref{eq:procova_lr_model}) has power $0.8$ for testing $H_0: \beta_1 = 0$ by considering the case of $W_{\mathrm{UN}}/W_{\text{P-LR}} \approx 1$ (so that $\zeta_{\mathrm{UN}} \approx \zeta_{\text{P-LR}}$), and the equation
\begin{align}
\label{eq:procova_lr_sample_size}
\frac{W_{\mathrm{UN}}}{W_{\text{P-LR}}} &\approx \frac{\widehat{\beta_1^*}\sqrt{N_{\mathrm{UN}}}/\hat{\beta}_1}{\sqrt{\mathrm{Var} \left ( \widehat{\beta_1^*} \right ) N_{\text{P-LR}} / \mathrm{Var} \left ( \hat{\beta}_1 \right )}} \nonumber \\
&\approx \sqrt{\frac{N_{\mathrm{UN}}}{N_{\text{P-LR}}} \left [ 1 - \frac{\mathrm{Var} \left ( \mu_{0,i} \right )}{E \left ( \mu_{0,i} \right ) \left \{1 - E \left ( \mu_{0,i} \right ) \right \}} \right ]} \nonumber \\
&= f_{\mathrm{EFF}} \sqrt{\frac{N_{\mathrm{UN}}}{N_{\text{P-LR}}}}.
\end{align}
Therefore, our sample size reduction formula for PROCOVA-LR compared to the unadjusted analysis is $N_{\text{P-LR}} = f_{\mathrm{EFF}}^2 N_{\mathrm{UN}}$. We can utilize prior point estimates or knowledge of the coefficients in the PROCOVA-LR model to prospectively estimate $\mathrm{Var} \left ( \mu_{0,i} \right )$ and $E \left ( \mu_{0,i} \right )$ in this equation. It is important to recognize that the magnitudes of $\Delta_{\mathrm{RD}}, \Delta_{\mathrm{RR}}$, and $\Delta_{\mathrm{OR}}$ are not involved in this calculation, and that this approach can be implemented based solely on historical information.

\begin{figure}[h]
\centering
\includegraphics[scale=0.35]{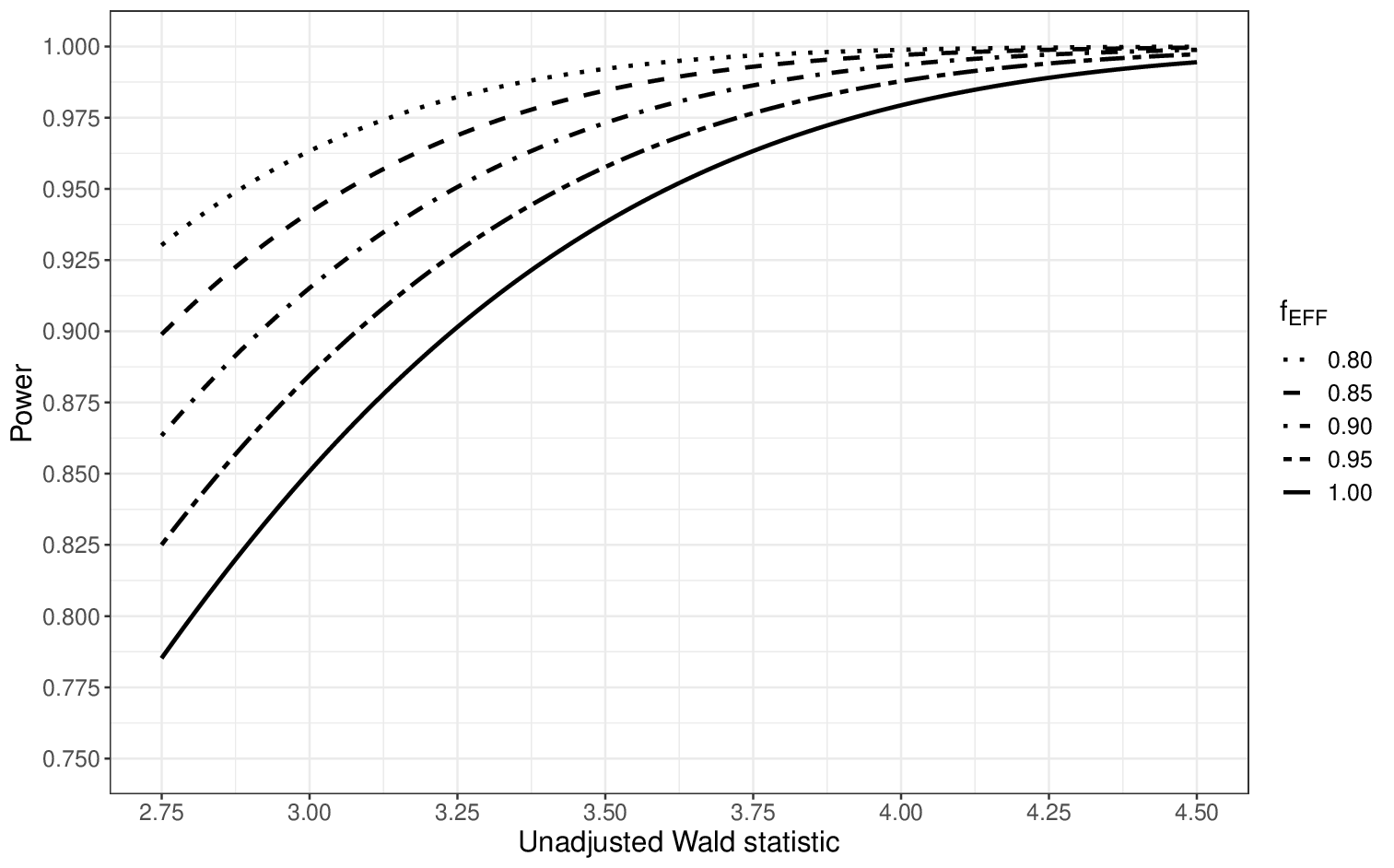}
\caption{Power for PROCOVA-LR as a function of the Wald test statistic $W_{\mathrm{UN}}$ for the unadjusted analysis and the efficiency factor $f_{\mathrm{EFF}}$. The power curve of the unadjusted analysis corresponds to $f_{\mathrm{EFF}} = 1$.}
\label{fig:factor_power}
\end{figure}

Our formula for the power gain of PROCOVA-LR compared to the unadjusted analysis is analogously derived by first approximating $W_{\text{P-LR}}$ as a function of $W_{\mathrm{UN}}$ and $f_{\mathrm{EFF}}$ based on equation (\ref{eq:efficiency_factor}), and then incorporating that approximation into the power calculation in equation (\ref{eq:power_adj}). More formally, for a fixed sample size $N$, we approximate $W_{\text{P-LR}} \approx W_{\mathrm{UN}}/f_{\mathrm{EFF}}$ and
\begin{equation}
\label{eq:procova_lr_power_gain}
\zeta_{\text{P-LR}} \approx \Phi \left (  \Phi^{-1} \left ( \frac{\alpha}{2} \right ) + \frac{W_{\mathrm{UN}}}{f_{\mathrm{EFF}}} \right ) + \Phi \left ( \Phi^{-1} \left ( \frac{\alpha}{2} \right ) - \frac{W_{\mathrm{UN}}}{f_{\mathrm{EFF}}} \right ).
\end{equation}
This approximation indicates that the power gain $\zeta_{\text{P-LR}} - \zeta_{\mathrm{UN}}$ depends both on the unadjusted Wald test statistic (alternatively, the unadjusted power $\zeta_{\mathrm{UN}}$) and $f_{\mathrm{EFF}}$. Figure \ref{fig:factor_power} visualizes the relationships between PROCOVA-LR power, $W_{\mathrm{UN}}$, and $f_{\mathrm{EFF}}$ for the cases of $f_{\mathrm{EFF}} = 0.8, 0.85, 0.9, 0.95, 1$. In this figure, the range of the $y$-axis corresponds to the power levels of interest in practice, and the power curve of the unadjusted analysis is obtained from $f_{\mathrm{EFF}} = 1$. 

\subsection{Inferring Risk Differences, Relative Risks, and Odds Ratios}
\label{sec:estimand}

In addition to considering sample size reductions and power gains under PROCOVA-LR with respect to the Wald test for the conditional odds ratio estimand, we calculate the variances and Wald test statistics for the g-computation based inferences on the marginal estimands $\Delta_{\mathrm{RD}}, \Delta_{\mathrm{RR}}$, and $\Delta_{\mathrm{OR}}$. The efficiency factor underlies the variance reduction of the estimators under PROCOVA-LR relative to those under the unadjusted analysis. This can be attributed to two connections between the logistic regression coefficients and the marginal estimands. First, the null hypothesis $H_0: \beta_1 = 0$ implies that $\Delta_{\mathrm{RD}} = 0, \Delta_{\mathrm{RR}} = 1$, and $\Delta_{\mathrm{OR}} = 1$. It is also for this reason that, in superiority trials, the power of traditional logistic regression and g-computation can be compared; they are testing equivalent null hypotheses of no treatment effects. Second, hypothesis tests for the marginal estimands are performed under PROCOVA-LR by utilizing g-computation with the coefficients in the PROCOVA-LR model (\ref{eq:procova_lr_model}), and the efficiency factor arises in the g-computation of the Wald test statistics for the coefficients. Ultimately, as demonstrated in Section \ref{sec:sample_size}, the efficiency factor is central to sample size calculations and power evaluations for the test of $H_0: \beta_1 = 0$ versus $H_0: \beta_1^* = 0$, and consequently it is an important consideration of tests for $H_0: \Delta_{\mathrm{RD}} = 0$, $H_0: \Delta_{\mathrm{RR}} = 1$, and $H_0: \Delta_{\mathrm{OR}} = 1$ when model (\ref{eq:procova_lr_model}) is the true data generating mechanism. 

The g-computation point estimator of a marginal estimand is obtained via a transformation $G: \mathbb{R}^3 \rightarrow \mathbb{R}$ of the estimators $\hat{\theta}_N = \left ( \hat{\beta}_0, \hat{\beta}_1, \hat{\beta}_2 \right )^{\mathsf{T}}$ from the fitted model (\ref{eq:procova_lr_model}). The calculation of the variances for $\widehat{\Delta_{\mathrm{RD}}}, \widehat{\Delta_{\mathrm{RR}}}$, and $\widehat{\Delta_{\mathrm{OR}}}$, and the derivations of the Wald test statistics for $\Delta_{\mathrm{RD}}, \Delta_{\mathrm{RR}}$, and $\Delta_{\mathrm{OR}}$, follow directly from the combination of the Delta method with g-computation from the fitted PROCOVA-LR model. Specifically, the Wald test statistic is defined using the transformation according to
\begin{equation*}
W = N^{1/2} G(\hat{\theta}_N) \{ J^{\mathsf{T}} \hat{V}_N J \}^{-1/2},
\end{equation*}
where $J$ is the $3 \times 1$ Jacobian associated with the transformation $G$. We demonstrate the calculations for the risk difference and the natural logarithm of the relative risk. The transformations for $\Delta_{\mathrm{RD}}$ and $\mathrm{log} \left ( \Delta_{\mathrm{RR}} \right )$ are
\begin{align*}
G_{\mathrm{RD}}(\hat{\beta}_0,\hat{\beta}_1,\hat{\beta}_2) &= \frac{1}{N}\sum_{i=1}^N \left \{ \frac{\mathrm{exp}(\hat{\beta}_0+\hat{\beta}_1+\hat{\beta}_2m_i)}
                                                        {1+\mathrm{exp}(\hat{\beta}_0+\hat{\beta}_1+\hat{\beta}_2m_i)} \right \} -
                                                        \frac{1}{N}\sum_{i=1}^N \left \{ \frac{\mathrm{exp}(\hat{\beta}_0+\hat{\beta}_2m_i)}
                                                        {1+\mathrm{exp}(\hat{\beta}_0+\hat{\beta}_2m_i)} \right \}, \\
G_{\mathrm{log} ( \mathrm{RR})}(\hat{\beta}_0,\hat{\beta}_1,\hat{\beta}_2) &= \mathrm{log} \left [ \frac{1}{N}\sum_{i=1}^N \left \{ \frac{\mathrm{exp}(\hat{\beta}_0+\hat{\beta}_1+\hat{\beta}_2m_i)}
                                                        {1+\mathrm{exp}(\hat{\beta}_0+\hat{\beta}_1+\hat{\beta}_2m_i)} \right \} \right ] -
                                                        \mathrm{log} \left [ \frac{1}{N}\sum_{i=1}^N \left \{ \frac{\mathrm{exp}(\hat{\beta}_0+\hat{\beta}_2m_i)}
                                                        {1+\mathrm{exp}(\hat{\beta}_0+\hat{\beta}_2m_i)} \right \} \right ].
\end{align*}
To simplify the notations for the Jacobians, we define 
\[
J_{01} = \displaystyle \frac{1}{N}\sum_{i=1}^N \left [ \frac{\mathrm{exp}(\hat{\beta}_0+\hat{\beta}_1+\hat{\beta}_2m_i)}{\{1+\mathrm{exp}(\hat{\beta}_0+\hat{\beta}_1+\hat{\beta}_2m_i)\}^2} \right ], \ J_{02} = \displaystyle \frac{1}{N}\sum_{i=1}^N \left [ \frac{\mathrm{exp}(\hat{\beta}_0+\hat{\beta}_2m_i)}{\{1+\mathrm{exp}(\hat{\beta}_0+\hat{\beta}_2m_i)\}^2} \right ],
\]
\[
J_{21} = \displaystyle \frac{1}{N}\sum_{i=1}^N \left [ \frac{m_i\mathrm{exp}(\hat{\beta}_0+\hat{\beta}_1+\hat{\beta}_2m_i)}{\{1+\mathrm{exp}(\hat{\beta}_0+\hat{\beta}_1+\hat{\beta}_2m_i)\}^2} \right ], J_{22} = \displaystyle \frac{1}{N}\sum_{i=1}^N \left [ \frac{m_i\mathrm{exp}(\hat{\beta}_0+\hat{\beta}_2m_i)}{\{1+\mathrm{exp}(\hat{\beta}_0+\hat{\beta}_2m_i)\}^2} \right ].
\]
The Jacobian for the first transformation is 
\begin{align*}
    J_{\mathrm{RD}} &= \begin{pmatrix}
        \frac{\partial G_{\mathrm{RD}}}{\partial \hat{\beta}_0}  \\
        \frac{\partial G_{\mathrm{RD}}}{\partial \hat{\beta}_1} \\
        \frac{\partial G_{\mathrm{RD}}}{\partial \hat{\beta}_2}
    \end{pmatrix} =
    \begin{pmatrix}
        J_{01} - J_{02} \\
        J_{01} \\
        J_{21} - J_{22}
    \end{pmatrix},
\end{align*}
and the Jacobian for the second transformation is  
\begin{align*}
    J_{\mathrm{log}(\mathrm{RR})} &= \begin{pmatrix}
        \frac{\partial G_{\mathrm{log}(\mathrm{RR})}}{\partial \hat{\beta}_0} \\
        \frac{\partial G_{\mathrm{log}(\mathrm{RR})}}{\partial \hat{\beta}_1} \\
        \frac{\partial G_{\mathrm{log}(\mathrm{RR})}}{\partial \hat{\beta}_2} 
    \end{pmatrix} =
    \begin{pmatrix}
        \displaystyle \left [ \frac{1}{N}\sum_{i=1}^N \left \{ \frac{\mathrm{exp}(\hat{\beta}_0+\hat{\beta}_1+\hat{\beta}_2m_i)}{1+\mathrm{exp}(\hat{\beta}_0+\hat{\beta}_1+\hat{\beta}_2m_i)} \right \} \right ]^{-1} J_{01} -
        \displaystyle \left [ \frac{1}{N}\sum_{i=1}^N \left \{ \frac{\mathrm{exp}(\hat{\beta}_0+\hat{\beta}_2m_i)}{1+\mathrm{exp}(\hat{\beta}_0+\hat{\beta}_2m_i)} \right \} \right ]^{-1} J_{02} \\
        \displaystyle  \left [ \frac{1}{N}\sum_{i=1}^N \left \{ \frac{\mathrm{exp}(\hat{\beta}_0+\hat{\beta}_1+\hat{\beta}_2m_i)}{1+\mathrm{exp}(\hat{\beta}_0+\hat{\beta}_1+\hat{\beta}_2m_i)} \right \} \right ]^{-1} J_{01} \\
        \displaystyle \left [ \frac{1}{N}\sum_{i=1}^N \left \{ \frac{\mathrm{exp}(\hat{\beta}_0+\hat{\beta}_1+\hat{\beta}_2m_i)}{1+\mathrm{exp}(\hat{\beta}_0+\hat{\beta}_1+\hat{\beta}_2m_i)} \right \} \right ]^{-1} J_{21} -
        \displaystyle \left [ \frac{1}{N}\sum_{i=1}^N \left \{ \frac{\mathrm{exp}(\hat{\beta}_0+\hat{\beta}_2m_i)}{1+\mathrm{exp}(\hat{\beta}_0+\hat{\beta}_2m_i)} \right \} \right ]^{-1} J_{22}
    \end{pmatrix}.    
\end{align*}
The transformations and Jacobians under the unadjusted model can be calculated in a similar manner as for the PROCOVA-LR model. Calculations for the natural logarithm of the odds ratio can also be performed, but we omit them here as the algebra becomes unwieldly. 

The Delta method can fail to yield an accurate variance estimator in practice, because it uses a first-order approximation of the variance whereas the transformations of the regression coefficients that are used to define the estimands are nonlinear. In addition, the Delta method fundamentally depends on the specification of the analysis model, because it utilizes the covariance matrix of the regression coefficient estimators. \citet{ye2023robust} effectively address these issues with their method to generate model-robust variance estimators using the sample covariance matrix. Their method uses augmented inverse probability weighting methods. However, a potential limitation of their method is that the sample covariance matrix may not yield efficient estimators in finite samples. We investigate the performance of the Delta method-based approach in Section \ref{sec:sim}.

In addition to the variances of the g-computation estimators and the Wald test statistics for the marginal estimands, we calculate confidence intervals for the marginal estimands using either the variance estimators from the Delta method or the nonparametric bootstrap. Although computationally more intensive, the nonparametric bootstrap is agnostic to whether the analysis model underlies the true data generating mechanism, and so we recommend the combination of g-computation with the nonparametric bootstrap to construct confidence intervals for the estimands. This corresponds with regulatory guidance on the use of the bootstrap for analyses that involve covariate adjustment \citep[p.~5]{food_and_drug_administration_adjusting_2023}. The simulation study for this approach to construct confidence intervals is described in Section \ref{sec:sim}.

It is important to note that regulatory guidance justifies testing on a parameter and reporting inferences on other, distinct estimands. For example, it is acceptable to regulatory agencies to perform a test on the hazard ratio and then report treatment effect inferences in terms of median survival. This further supports our approach under PROCOVA-LR to perform both hypothesis tests on the conditional odds ratio estimand and additional inferences on the marginal estimands.

\subsection{Cases of Model Misspecification}
\label{sec:misspecified}

The validity of the inferences for the different estimands under PROCOVA-LR, as well as of the efficiency factor in equation (\ref{eq:efficiency_factor}), can be affected as a result of model misspecification. Three types of model misspecifications are common in practice: the omission of an important covariate, a shift in the prognostic scores, and random errors in the prognostic scores. We demonstrate that the efficiency factor is valid under the first two types of model misspecifications. For the third type, we propose an adjustment to the efficiency factor that can yield accurate prospective predictions of the gains that can result from PROCOVA-LR. The accuracy of this adjustment for $f_{\mathrm{EFF}}$ is demonstrated via simulation studies in Section \ref{sec:sim}.

We first consider the omission of an important covariate from both the procedure for constructing the prognostic scores, as well as from direct adjustment in the logistic regression model. PROCOVA-LR then provides only a partial adjustment for the covariates in this case, as the omitted covariate is neither contained in the prognostic score nor as a predictor variable in the model. \citet{gail1988tests} demonstrated that, for practical purposes, logistic regression models that utilize only a partial adjustment yield valid tests for the null hypothesis of the treatment effect. In addition, the efficiency factor in equation (\ref{eq:efficiency_factor}) remains valid because \citet{neuhaus1998estimation} used robust estimates of variance in his derivation.

Next, we consider the case in which the prognostic score $m_i$ in the PROCOVA-LR analysis is not the true predictor underlying the data generation mechanism, but that instead a shifted version $\widetilde{m}_i = m_i + b$ of it is the true predictor for data generation. The shift is defined according to the bias term $b$. As the PROCOVA-LR model has an intercept term, parameter $\beta_0$ absorbs bias $b$. Hence, the PROCOVA-LR analyses remain valid. In addition, the calculation of the efficiency factor based on the $m_i$ will be similar to the true efficiency factor that would have been calculated if the $\widetilde{m}_i$ were observable. This is because, although the $m_i$ and $\widetilde{m}_i$ differ, the corresponding values for the participants' probabilities of an event under control, i.e., the $\mu_{0,i}$, will not differ as much after the logistic transformation. Hence, the $E(\mu_{0,i})$ will be fairly similar when calculated using either $m_i$ or $\widetilde{m}_i$. As the bias term is additive, the variances of the $m_i$ and $\widetilde{m}_i$ will also be the same, and so too will the corresponding values of $\mathrm{Var}(\mu_{0,i})$.

The third case is a more general situation compared to the second case, in which the observed prognostic scores $m_i$ differ from the true prognostic scores $\widetilde{m}_i$ by random error terms, i.e., $\widetilde{m}_i = m_i + \delta_i$ for random variables $\delta_i$. This case also corresponds to logistic regression with errors in variables, which has been studied by \citet{stefanski1985covariate} and \citet{huang2001consistent}. Previous investigations in this domain have not considered the validity of statistical tests for the coefficients, but instead primarily focused on adjusting the MLEs so that they are asymptotically unbiased. We demonstrate via simulation studies in Section \ref{sec:sim_misspecified} that the Wald test maintains its nominal significance level in this case. The efficiency factor in equation (\ref{eq:efficiency_factor}) is not directly applicable for the case of random errors in the prognostic scores. This is because the efficiency factor involves the variance of the $\mu_{0,i}$, and the observed prognostic scores $m_i$ that are used in the PROCOVA-LR analysis to estimate this variance will contain spurious variability. Hence, $\mathrm{Var}(\mu_{0,i})$ will be over-estimated, and the efficiency factor will overestimate the benefit of adjustment by the prognostic score in PROCOVA-LR. To address this overestimation and more accurately estimate the gains of PROCOVA-LR in this case, we propose to adjust (\ref{eq:efficiency_factor}) by using the fact that the squared correlation between the $\mu_{0,i}$ that are calculated based on $m_i$ and the $\widetilde{\mu}_{0,i}$ that are calculated based on the $\widetilde{m}_i$ corresponds to the percentage of variance in $\widetilde{\mu}_{0,i}$ that can be explained by $\mu_{0,i}$. Hence, we adjust the efficiency factor from equation (\ref{eq:efficiency_factor}) by this correlation according to 
\begin{equation}
\label{eq:factor_adj}
\tilde{f}_{\mathrm{EFF}} = \sqrt{1-\frac{\mathrm{Var}(\mu_{0,i})\mathrm{Corr}(\tilde{\mu}_{0,i},\mu_{0,i})^2}{E(\mu_{0,i}) \left \{ 1-E(\mu_{0,i}) \right \}}}.
\end{equation}
to remedy the risk of overconfidence in PROCOVA-LR in the case of random errors in the prognostic scores. We demonstrate the efficacy of equation (\ref{eq:factor_adj}) via simulation studies in Section \ref{sec:sim_misspecified}. The correlation between $\mu_{0,i}$ and $\widetilde{\mu}_{0,i}$ is unknown in practice, and one straightforward approach to estimate this correlation is by using the concordance index between the observed and predicted binary outcomes.

\section{Simulation Studies}
\label{sec:sim}

\subsection{Data Generation Mechanisms and Evaluation Metrics}
\label{sec:outline_of_settings}

We design two sets of simulation studies to investigate the frequentist properties of PROCOVA-LR compared to the unadjusted analysis across several data generation mechanisms. These comparisons are performed in terms of inferences for the coefficient $\beta_1$ associated with the treatment indicator, and g-computation inferences for $\Delta_{\mathrm{RD}}, \Delta_{\mathrm{RR}}$, and $\Delta_{\mathrm{OR}}$. The metrics for the simulation studies are $E \left ( \mu_{0,i} \right )$, $\mathrm{Var} \left ( \mu_{0,i} \right )$, $f_{\mathrm{EFF}}$, the Type I error rate control and power of the tests for the estimands, the biases of point estimators for the estimands, and the expected widths of $95\%$ confidence intervals for the estimands. The analyses were implemented using the \texttt{glm} function in \texttt{R}. In our evaluations for $\Delta_{\mathrm{RD}}, \Delta_{\mathrm{RR}}$, and $\Delta_{\mathrm{OR}}$ for each scenario, we set the true value of each estimand as the average of the estimands across the simulated datasets. We utilize $5000$ nonparametric bootstrap samples for each simulated dataset to construct the confidence intervals via the percentile method.

The first set of simulation studies consists of four scenarios in which the true data generation mechanism is the PROCOVA-LR model (\ref{eq:procova_lr_model}). Each mechanism is defined by a distribution on the prognostic scores $m_i$, values of the parameters $\beta_0, \beta_1$, and $\beta_2$ from equation (\ref{eq:procova_lr_model}), and a RCT sample size $N$. The scenarios are summarized in Table \ref{tab:simulation_studies_1_mechanisms}. To simplify the design of both sets of simulation studies, we utilize Normal distributions to simulate prognostic scores and other covariates that are used in the generation of the outcomes and PROCOVA-LR analysis model. This choice corresponds in practice to transforming the original predictors, e.g., by centering them or applying the logit transformation to the prognostic score that was originally defined as the predicted probability of an event from the digital twin distribution. Prognostic scores are independent and identically distributed according to their respective distributions. The first two scenarios are considered so as to illustrate that adjustment by the prognostic score improves the efficiency of inferences for the estimands, and that the magnitude of $\beta_1$ does not affect the efficiency gains. The third and fourth scenarios demonstrate the effects of the variance of the prognostic scores $\mathrm{Var} \left ( \mu_{0,i} \right )$ and the underlying prevalence of the event under control $E \left ( \mu_{0,i} \right )$, respectively, on the efficiency gains of PROCOVA-LR. The randomization ratio for all RCTs is 1:1. We simulate $10^5$ RCTs for each scenario to control the Monte Carlo errors of the metrics. The Type I error rates were calculated based on $10^5$ simulated datasets with $\beta_1 = 0$. The results of these simulation studies are in Section \ref{sec:sim_correct_model}.

\begin{table}[H]
\centering
\resizebox{\columnwidth}{!}{
\begin{tabular}{|l|c|c|c|}
\hline
Scenario & Distribution of $m_i$ & Data Generation Mechanism & $N$ \\ 
\hline \hline
Baseline & $\mathrm{Normal} \left (0, 1.5^2 \right )$ & $\mathrm{logit} \left \{ \mathrm{Pr} \left ( y_i = 1 \mid w_i, m_i \right ) \right \} = 1 + 0.75w_i + m_i$ & $500$ \\
\hline 
Large Effect & $\mathrm{Normal} \left (0, 1.5^2 \right )$ & $\mathrm{logit} \left \{ \mathrm{Pr} \left ( y_i = 1 \mid w_i, m_i \right ) \right \} = 1 + 0.85w_i + m_i$ & $500$\\
\hline
Large Variance & $\mathrm{Normal} \left (0, 2.5^2 \right )$ & $\mathrm{logit} \left \{ \mathrm{Pr} \left ( y_i = 1 \mid w_i, m_i \right ) \right \} = 1 + 0.75w_i + m_i$ & $500$ \\
\hline
High Prevalence & $\mathrm{Normal} \left (0, 2^2 \right )$ & $\mathrm{logit} \left \{ \mathrm{Pr} \left ( y_i = 1 \mid w_i, m_i \right ) \right \} = 2.5 + 0.75w_i + m_i$ & $800$ \\
\hline
\end{tabular}
}
\caption{Data generation mechanisms for four scenarios in which the data are generated according to the PROCOVA-LR model.}
\label{tab:simulation_studies_1_mechanisms}
\end{table}

\normalsize

The second set of simulation studies consists of three common scenarios that can arise in practice in which discrepancies exist between the true data generation mechanism and the PROCOVA-LR model that is used to analyze the data. The scenarios are summarized in Table \ref{tab:simulation_studies_2_mechanisms}. We explore the properties of both the (misspecified) unadjusted and PROCOVA-LR model across these scenarios. In the first scenario, an important covariate $x_i \in \mathbb{R}$ is involved in the data generation along with the prognostic score $m_i$, but $x_i$ is omitted from the analyses. The $x_i$ and $m_i$ have a correlation of $4/9$ in this scenario. The second scenario considers the case in which $m_i$ is a noisy version of $x_i$, with $m_i = x_i + \delta_i$ for independent $\delta_i \sim \mathrm{Normal} \left ( 0, 1^2 \right )$, and $x_i$ is the true driver of data generation. The third scenario extends the consideration from the second scenario in that $m_i$ is a noisy and shifted version of $x_i$, with $m_i = x_i + 0.5 + \delta_i$ for independent $\delta_i \sim \mathrm{Normal} \left ( 0, 1^2 \right )$. As before, we simulate $10^5$ RCTs, each with a 1:1 randomization ratio, for each scenario. The results are summarized in Section \ref{sec:sim_misspecified}.

\begin{table}[H]
\centering
\resizebox{\columnwidth}{!}{
\begin{tabular}{|l|c|c|c|}
\hline
Scenario & Distribution of Covariates & Data Generation Mechanism & $N$ \\ 
\hline \hline
Omitted Covariate & $\begin{pmatrix} m_i \\ x_i \end{pmatrix} \sim \mathrm{Normal} \left ( \begin{pmatrix} 0 \\ 0 \end{pmatrix}, \begin{pmatrix} 2.25 & 1 \\ 1 & 2.25 \end{pmatrix} \right ) $ & $\mathrm{logit} \left \{ \mathrm{Pr} \left ( y_i = 1 \mid w_i, m_i, x_i \right ) \right \} = 0.75w_i + m_i + x_i$ & $800$ \\
\hline 
Random Error & $\begin{pmatrix} m_i \\ x_i \end{pmatrix} \sim \mathrm{Normal} \left ( \begin{pmatrix} 1 \\ 1 \end{pmatrix}, \begin{pmatrix}  3.25 & 2.25 \\ 2.25 & 2.25 \end{pmatrix}\right ) $ & $\mathrm{logit} \left \{ \mathrm{Pr} \left ( y_i = 1 \mid w_i, m_i, x_i \right ) \right \} = 0.75w_i + x_i$ & $500$\\
\hline
Shift and Random Error & $\begin{pmatrix} m_i \\ x_i \end{pmatrix} \sim \mathrm{Normal} \left ( \begin{pmatrix} 1.5 \\ 1 \end{pmatrix}, \begin{pmatrix} 3.25 & 2.25 \\ 2.25 & 2.25 \end{pmatrix}\right ) $ & $\mathrm{logit} \left \{ \mathrm{Pr} \left ( y_i = 1 \mid w_i, m_i, x_i \right ) \right \} = 0.75w_i + x_i$ & $500$ \\
\hline
\end{tabular}
}
\caption{Three scenarios in which discrepancies exist between the true data generation mechanism and the PROCOVA-LR model that is used to analyze the data.}
\label{tab:simulation_studies_2_mechanisms}
\end{table}

\subsection{Cases With a Correctly Specified PROCOVA-LR Model}
\label{sec:sim_correct_model}

\normalsize

The results of the first set of simulation studies for the scenarios from Table \ref{tab:simulation_studies_1_mechanisms} are summarized in Tables \ref{tab:simulation_studies_1_power}, \ref{tab:simulation_studies_1_Type_I_error_rate}, and \ref{tab:simulation_studies_1_g_computation_inferences}. The first two of these tables compare power and Type I error rates across the scenarios, and illustrate the connection between the efficiency factor and the power gain of PROCOVA-LR over the unadjusted analysis on conditional and marginal estimands. Table \ref{tab:simulation_studies_1_g_computation_inferences} quantifies the asymptotic unbiasedness and efficiency of PROCOVA-LR compared to the unadjusted analysis with respect to the g-computation inferences on $\Delta_{\mathrm{RD}}, \Delta_{\mathrm{RR}}$, and $\Delta_{\mathrm{OR}}$. 

In Table \ref{tab:simulation_studies_1_power}, we record for each scenario the averages and standard deviations of the ratio of the Wald test statistics $W_{\mathrm{UN}}/W_{\text{P-LR}}$ for the unadjusted versus PROCOVA-LR analyses, the $E \left ( \mu_{0,i} \right )$, and the efficiency factors $f_{\mathrm{EFF}}$ across the simulated datasets. The standard deviations are in parentheses directly below the averages. We also record the averages of the $\mathrm{Var} \left ( \mu_{0,i} \right )$ across the simulated datasets for each scenario. We observe that the averages of $E \left ( \mu_{0,i} \right )$ and $\mathrm{Var} \left ( \mu_{0,i} \right )$ are related to the expectations of $f_{\mathrm{EFF}}$ according to equation (\ref{eq:efficiency_factor}). In addition, the averages of the $W_{\mathrm{UN}}/W_{\text{P-LR}}$ correspond to the expectations of the $f_{\mathrm{EFF}}$ across the scenarios. Each realized ratio of Wald test statistics can be interpreted as a realized efficiency factor. The standard deviations of the $E \left ( \mu_{0,i} \right )$ and $f_{\mathrm{EFF}}$ are consistently small across all scenarios, but the standard deviations of the $W_{\mathrm{UN}}/W_{\text{P-LR}}$ can be large because $\hat{\beta}_1$ and $\widehat{\beta_1^*}$ could be small in absolute value for some simulated datasets. The power gains of PROCOVA-LR compared to the unadjusted analysis as evaluated via these simulations are observed to be related to the $W_{\mathrm{UN}}/W_{\text{P-LR}}$ and $f_{\mathrm{EFF}}$ as in equation (\ref{eq:procova_lr_power_gain}). These results indicate that PROCOVA-LR can increase the power of significance tests for $\beta_1, \Delta_{\mathrm{RD}}$, and $\mathrm{log} \left ( \Delta_{\mathrm{RR}} \right )$.

Comparing the results for the Baseline and Large Effect scenarios in Table \ref{tab:simulation_studies_1_power}, we observe that the magnitude of $\beta_1$ does not substantively affect the power gains of PROCOVA-LR. The Large Variance scenario indicates how a larger $\mathrm{Var} \left ( \mu_{0,i} \right )$ corresponds to a smaller expected $f_{\mathrm{EFF}}$ and $W_{\mathrm{UN}}/W_{\text{P-LR}}$, as well as reduced power for PROCOVA-LR compared to the Baseline scenario. However, the Large Variance scenario also indicates a larger power gain of PROCOVA-LR compared to the unadjusted analysis. Similarly, the High Prevalence scenario demonstrates that the benefits of PROCOVA-LR decrease as $E \left ( \mu_{0,i} \right ) \rightarrow 1$ compared to the scenarios in which the prevalence is more moderate, and that the PROCOVA-LR exhibits a larger power gain over the unadjusted analysis compared to the Baseline scenario. In practice, $E \left ( \mu_{0,i} \right ) \rightarrow 0$ or $E \left ( \mu_{0,i} \right ) \rightarrow 1$ correspond to rare or commonplace events, respectively, with $\mathrm{Var} \left ( \mu_{0,i} \right ) \rightarrow 0$ in either case, so that covariate adjustment won't be expected \emph{a priori} to yield improved treatment effect inferences. All of these results further demonstrate the validity of the theory of \citet{neuhaus1998estimation}, and its applicability via the efficiency factor in equation (\ref{eq:efficiency_factor}) for power gain of the Wald test under PROCOVA-LR. Our results also match the empirical findings of \citet{hernandez2004covariate}, who apparently were not aware that the theory of \citet{neuhaus1998estimation} could justify their findings. Supplementary results demonstrating additional correspondences between the bias factor, AREs, and efficiency factors for all scenarios are in Table \ref{tab:app_bias_factor} in Appendix \ref{app:additional_simulation}. 

\begin{table}[h]
\centering
\resizebox{\columnwidth}{!}{
\begin{tabular}{|l|c|c|c|c|c|c|c|c|c|c|c|c|}
\hline
& \multicolumn{3}{c|}{Efficiency Factor} & \multicolumn{3}{c|}{Test on $\beta_1$} & \multicolumn{3}{c|}{Test on $\Delta_{\mathrm{RD}}$} & \multicolumn{3}{c|}{Test on $\mathrm{log} \left ( \Delta_{\mathrm{RR}} \right )$} \\
& \multicolumn{3}{c|}{Elements} & \multicolumn{1}{c|}{Wald Ratio} & \multicolumn{2}{c|}{Power} & \multicolumn{1}{c|}{Wald Ratio} & \multicolumn{2}{c|}{Power} & \multicolumn{1}{c|}{Wald Ratio} & \multicolumn{2}{c|}{Power} \\
\hline
Scenario  & $E(\mu_{0,i})$  & $\mathrm{Var}(\mu_{0,i})$  & $f_{\mathrm{EFF}}$  & $W_{\mathrm{UN}}/W_{\text{P-LR}}$ & UN & P-LR & $W_{\mathrm{UN}}/W_{\text{P-LR}}$ & UN & P-LR & $W_{\mathrm{UN}}/W_{\text{P-LR}}$ & UN & P-LR \\
\hline \hline
Baseline & 0.67 & 0.06 & 0.85 & 0.83 & 77.9 & 89.0 & 0.82 & 78.1 & 89.3 & 0.82 & 77.7 & 89.1 \\
     & (0.01) & - & (0.01) & (11.8) & & & (11.8) & & & (11.8) & & \\
\hline
Large & 0.67 & 0.06 & 0.85 & 0.87 & 86.6 & 94.6 & 0.86 & 86.8 & 94.8 & 0.86 & 86.5 & 94.7 \\
Effect & (0.01) & - & (0.01) & (0.27) & & & (0.27) & & & (0.27) & & \\
\hline
Large & 0.63 & 0.11 & 0.73 & 0.75 & 55.5 & 81.5 & 0.74 & 55.7 & 82.0 & 0.74 & 55.3 & 81.8 \\
Variance & (0.01) & - & (0.01) & (2.15) & & & (2.16) & & & (2.15) & & \\
\hline
High & 0.83 & 0.05 & 0.82 & 0.83 & 70.5 & 85.4 & 0.83 & 71.1 & 85.8 & 0.82 & 70.7 & 85.7 \\
Prevalence & (0.01) & - & (0.01) & (0.50) & & & (0.50) & & & (0.50) & & \\
\hline
\end{tabular}
}
\caption{Results on efficiency factors and power gains from the simulation studies involving the scenarios in Table \ref{tab:simulation_studies_1_mechanisms}. PROCOVA-LR (abbreviated as ``P-LR'') exhibits increased power for significance testing of $\beta_1$, $\Delta_{\mathrm{RD}}$, and $\mathrm{log} \left ( \Delta_{\mathrm{RR}} \right )$. The power gain of PROCOVA-LR compared to the unadjusted analysis (abbreviated as ``UN'') follows the prediction based on the efficiency factor and equation (\ref{eq:procova_lr_power_gain}).}
\label{tab:simulation_studies_1_power}
\end{table}

\normalsize

Type I error rates for the Baseline, Large Variance, and High Prevalence scenarios are in Table \ref{tab:simulation_studies_1_Type_I_error_rate}, and correspond to the rejection rates of the Wald test when $\beta_1 = 0$ (and, by implication, $\Delta_{\mathrm{RD}} = 0$ and $\mathrm{log} \left ( \Delta_{\mathrm{RR}} \right ) = 0$). We observe that the Type I error rates for the tests on $\beta_1$ are controlled at the $\alpha = 0.05$ level, but the rejection rates for the tests on $\Delta_{\mathrm{RD}}$ and $\mathrm{log} \left ( \Delta_{\mathrm{RR}} \right )$ deviate from this nominal level. These discrepancies can be explained by the shortcoming of the Delta method in terms of its accuracy in estimating the variance of the point estimator. They should decrease in absolute value as the sample size increases. Besides the Delta method, tests can be performed by combining the nonparametric bootstrap with g-computation. Alternatively, confidence intervals constructed by means of the nonparametric bootstrap and g-computation can lead to tests with controlled Type I error rates.

\begin{table}[h]
\centering
\begin{tabular}{|l|c|c|c|c|c|c|}
\hline
 & \multicolumn{2}{c|}{Test on $\beta_1$} & \multicolumn{2}{c|}{Test on $\Delta_{\mathrm{RD}}$ } & \multicolumn{2}{c|}{Test on $\mathrm{log} \left ( \Delta_{\mathrm{RR}} \right )$} \\
 \hline
Scenario & \multicolumn{1}{c|}{UN} & \multicolumn{1}{c|}{P-LR} & \multicolumn{1}{c|}{UN} & \multicolumn{1}{c|}{P-LR} & \multicolumn{1}{c|}{UN} & \multicolumn{1}{c|}{P-LR} \\
\hline
\hline
Baseline & 5.02 & 5.06 & 5.10 & 5.25 & 4.96 & 5.12 \\
\hline
Large Variance & 4.92 & 4.96 & 5.08 & 5.21 & 4.85 & 5.10 \\
\hline 
High Prevalence & 4.82 & 4.99 & 4.96 & 5.19 & 4.85 & 5.10 \\
\hline
\end{tabular}
\caption{Type I error rates for the Wald tests on $\beta_1, \Delta_{\mathrm{RD}}$, and $\mathrm{log} \left ( \Delta_{\mathrm{RR}} \right )$. PROCOVA-LR (abbreviated as ``P-LR'') controls the Type I error rate for the test on $\beta_1$, but rejects more often than desired for the tests on $\Delta_{\mathrm{RD}}$ and $\mathrm{log} \left ( \Delta_{\mathrm{RR}} \right )$. In contrast, the unadjusted analysis (abbreviated as ``UN'') controls the Type I error rates for all tests.}
\label{tab:simulation_studies_1_Type_I_error_rate}
\end{table}

The summary of the frequentist properties of the g-computation inferences for $\Delta_{\mathrm{RD}}, \Delta_{\mathrm{RR}}$, and $\Delta_{\mathrm{OR}}$ under the Baseline, Large Variance, and High Prevalence scenarios in Table \ref{tab:simulation_studies_1_g_computation_inferences} indicates the consistency of the treatment effect estimators from the unadjusted and PROCOVA-LR analyses, and corresponds to the theory of \citet{freedman2008randomization}. It is important to recognize that, for each scenario and estimand, the true estimand value was obtained as the average of the finite-population estimand values across the simulated datasets. The Monte Carlo error associated with each such value (indicated in parentheses below the value) is two orders of magnitude less than the average. The standard deviations of the estimators and the average widths of the intervals under PROCOVA-LR are consistently smaller than those of the unadjusted analysis. The ratio of the average widths of the confidence intervals is approximately equal to the expected efficiency factor. By comparing the Baseline and Large Effect scenarios, we observe that the efficiency gain from covariate adjustment is not related to the magnitude of $\beta_1$.

\begin{table}[h]
    \centering
    \resizebox{\columnwidth}{!}{
    \begin{tabular}{|l|c|c|c|c|c|c|c|c|}
        \hline
         & Estimand & Estimand & \multicolumn{2}{c|}{Deviations of} & \multicolumn{3}{c|}{Width of} & $f_{\mathrm{EFF}}$ \\
         & & Value & \multicolumn{2}{c|}{Estimators} & \multicolumn{3}{c|}{$95\%$ CI} & \\
         \hline
         Scenario & & & UN & P-LR & UN & P-LR & Ratio & \\
        \hline
        \hline
        Baseline & $\Delta_{\mathrm{OR}}$ & 1.73 & -.047 & -.042 & 1.49 & 1.26 & .86 & .85\\ 
         & & (.02) & (.404) & (.352) & (.38) & (.29) & (.10) & (0.01) \\ 
         & $\Delta_{\mathrm{RR}}$ & 1.16 & -.002 & -.003 & .26 & .22 & .85 & .85 \\ 
         & & (.01) & (.070) & (.061) & (.02) & (.02) & (.04) & (0.01) \\ 
         & $\Delta_{\mathrm{RD}}$ & .109 & .0002 & -.0006 & .155 & .133 & .86 & .85 \\
         & & (.003) & (.0430) & (.0374) & (.004) & (.005) & (.03) & (0.01) \\ 
         \hline
        Large & $\Delta_{\mathrm{OR}}$ & 1.87 & -.072 & -.054 & 1.66 & 1.40 & .85 & .85\\
        Effect & & (.02) & (.404) & (.348) & (.39) & (.29) & (.09) & (0.01)\\
         & $\Delta_{\mathrm{RR}}$ & 1.18 & -.006 & -.005 & .26 & .22 & .85 & .85 \\
         & & (.01) & (.064) & (.056) & (.02) & (.02) & (.04) & (0.01)\\
         & $\Delta_{\mathrm{RD}}$ & .121 & -.0024 & -.0018 & .154 & .132 & .86 & .85 \\
         & & (.003) & (-.0387) & (.0336) & (.004) & (.005) & (.02) & (0.01)\\
        \hline
    \end{tabular}}
    \caption{Summary of the frequentist properties for the g-computation based point estimators and confidence intervals for the marginal estimands. The standard deviations of the mean values are indicated in parentheses below the means. The small values for the average deviances correspond to the consistency of the point estimators. The widths of the confidence intervals from PROCOVA-LR are shorter than those from the unadjusted analysis. The efficiency gains of PROCOVA-LR, in terms of both variance reductions of the point estimators and confidence interval width reductions, are closely aligned with the efficiency factor values.}
\label{tab:simulation_studies_1_g_computation_inferences}
\end{table}

\subsection{Cases With a Misspecified PROCOVA-LR Model}
\label{sec:sim_misspecified}

The results for the scenarios involving misspecified PROCOVA-LR models with respect to power gain, Type I error rate control, bias, and confidence interval width are summarized in Tables \ref{tab:simulation_studies_2_power}, \ref{tab:simulation_studies_2_Type_I_error_rate}, and \ref{tab:simulation_studies_2_g_computation_inferences}. In the Omitted Covariate scenario the efficiency factor is calculated using equation (\ref{eq:efficiency_factor}) as before, where $\mu_{0,i}$ is calculated based solely on $m_i$ without consideration of the omitted covariate $x_i$. In the last two scenarios from Table \ref{tab:simulation_studies_2_mechanisms}, the expected efficiency factors are calculated based on equation (\ref{eq:factor_adj}) and a known value for $\mathrm{Corr}(\tilde{\mu}_{0,i}, \mu_{0,i})$.

We observe from Table \ref{tab:simulation_studies_2_power} that, as before, the average of $E \left ( \mu_{0,i} \right )$ and $\mathrm{Var} \left ( \mu_{0,i} \right )$ are related to the expectations of $f_{\mathrm{EFF}}$ according to equation (\ref{eq:efficiency_factor}) (for the Omitted Covariate scenario) and equation (\ref{eq:factor_adj}) (for the latter two scenarios in Table \ref{tab:simulation_studies_2_mechanisms}), and that the averages of the $W_{\mathrm{UN}}/W_{\text{P-LR}}$ correspond to the expectations of $f_{\mathrm{EFF}}$. The omission of a covariate does not affect the validity of equation (\ref{eq:efficiency_factor}), and the performance in the Omitted Covariate case is similar to the Baseline case. The similarity in the expectation of $f_{\mathrm{EFF}}$ and the $W_{\mathrm{UN}}/W_{\text{P-LR}}$ in both the Random Error and the Shift and Random Error scenarios indicates the utility of equation (\ref{eq:factor_adj}). Furthermore, the performances of these two scenarios are observed to be similar, which can be explained by the systematic shift in the covariate being effectively absorbed by $\beta_0$. The range of $0.9$ to $0.92$ for the $W_{\mathrm{UN}}/W_{\text{P-LR}}$ in the Random Error and Shift and Random Error scenarios correspond to a sample size reduction of approximately $8\%$. The Random Error scenario has the same data generation mechanism as the Baseline scenario, but the additional error in the covariate leads to the $f_{\mathrm{EFF}}$ being closer to $1$, and a corresponding decrease in performance. Both of these scenarios are extreme in that the variance of the random error is $44 \%$ of the variance of the prognostic score. Our consideration of these extreme situations further demonstrates that efficiency gains are feasible for misspecified models. Ultimately, we conclude that PROCOVA-LR can improve the efficiency of treatment effect inferences compared to the unadjusted analysis even when the PROCOVA-LR model is misspecified, and that a key point is to obtain a good estimate of the correlation between $\tilde{\mu}_{0,i}$ and $\mu_{0,i}$ for the expected efficiency factor from equation (\ref{eq:factor_adj}) to be accurate. 

\begin{table}[h]
\centering
\resizebox{\columnwidth}{!}{
\begin{tabular}{|l|c|c|c|c|c|c|c|c|c|c|c|c|}
\hline
& \multicolumn{3}{c|}{Efficiency Factor} & \multicolumn{3}{c|}{Test on $\beta_1$} & \multicolumn{3}{c|}{Test on $\Delta_{\mathrm{RD}}$} & \multicolumn{3}{c|}{Test on $\mathrm{log} \left ( \Delta_{\mathrm{RR}} \right )$} \\
& \multicolumn{3}{c|}{Elements} & \multicolumn{1}{c|}{Wald Ratio} & \multicolumn{2}{c|}{Power} & \multicolumn{1}{c|}{Wald Ratio} & \multicolumn{2}{c|}{Power} & \multicolumn{1}{c|}{Wald Ratio} & \multicolumn{2}{c|}{Power} \\
\hline
Scenario  & $E(\mu_{0,i})$  & $\mathrm{Var}(\mu_{0,i})$  & $f_{\mathrm{EFF}}$  & $W_{\mathrm{UN}}/W_{\text{P-LR}}$ & UN & P-LR & $W_{\mathrm{UN}}/W_{\text{P-LR}}$ & UN & P-LR & $W_{\mathrm{UN}}/W_{\text{P-LR}}$ & UN & P-LR \\
\hline \hline
Omitted & 0.50 & 0.07 & 0.84$^{\dag}$ & 0.83 & 78.1 & 91.6 & 0.82 & 78.1 & 91.7 & 0.82 & 78.1 & 91.6 \\
Covariate & (0.01) & - & (0.01) & (0.63) & & & (0.64) & & & (0.63) & & \\
\hline 
Random & 0.67 & 0.06 & 0.90$^{\dag\dag}$ & 0.92 & 77.8 & 85.2 & 0.91 & 78.1 & 85.6 & 0.91 & 77.7 & 85.3 \\
Error & (0.01) & - & (0.01) & (3.43) & & & (3.43) & & & (3.43) & & \\
\hline  
Shift and & 0.67 & 0.06 & 0.90$^{\dag\dag}$ & 0.91 & 77.8 & 85.3 & 0.90 & 78.1 & 85.7 & 0.90 & 77.7 & 85.3 \\
Random Error & (0.01) & - & (0.01) & (0.52) & & & (0.52) & & & (0.52) & & \\
\hline
\end{tabular}
}
\caption{Results on efficiency factors and power gains for the scenarios from Table \ref{tab:simulation_studies_2_mechanisms}. PROCOVA-LR (abbreviated by ``P-LR'') exhibits increased power for the three types of hypothesis tests, and its power gain compared to the unadjusted analysis (abbreviated by ``UN'') follows the prediction based on the efficiency factor. $\dag$ The efficiency factor for this case is calculated using $m_i$ only, omitting $x_i$. $\dag\dag$ The expected efficiency factors for these two cases are calculated based on the adjusted efficiency factor in equation (\ref{eq:factor_adj}).}
\label{tab:simulation_studies_2_power}
\end{table}

\normalsize

The summary of the Type I error rates in Table \ref{tab:simulation_studies_2_Type_I_error_rate} demonstrates that both the unadjusted and PROCOVA-LR analyses typically reject the null hypothesis more often than desired. However, for inference on the conditional estimand, PROCOVA-LR maintains better control of the Type I error rate compared to the unadjusted analysis for the Omitted Covariate scenario.

\begin{table}[h]
    \centering
    \begin{tabular}{|c | c | c | c | c | c | c|}
        \hline
        & \multicolumn{2}{c|}{Test on $\beta_1$} & \multicolumn{2}{c|}{Test on $\Delta_{\mathrm{RD}}$} & \multicolumn{2}{c|}{Test on $\mathrm{log} \left ( \Delta_{\mathrm{RR}} \right )$} \\
        \hline
        Scenario & UN & P-LR & UN & P-LR & UN & P-LR \\
        \hline \hline
        Omitted Covariate & 5.17 & 5.02 & 5.17 & 5.13 & 5.17 & 5.05 \\
        \hline
        Random Error & 5.04 & 4.98 & 5.11 & 5.16 & 4.98 & 5.04 \\
        \hline
        Shift and Random Error & 4.99 & 4.94 & 5.08 & 5.13 & 4.91 & 4.99 \\
        \hline
    \end{tabular}
    \caption{Type I error rates for the Wald tests on $\beta_1$, $\Delta_{\mathrm{RD}}$, and $\mathrm{log} \left ( \Delta_{\mathrm{RR}} \right )$. Both PROCOVA-LR (abbreviated as ``P-LR'') and the unadjusted analysis (abbreviated as ``UN'') typically reject the null hypothesis more often than desired for the marginal estimands. PROCOVA-LR better controls the Type I error rate for the test on the conditional estimand $\beta_1$.}
    \label{tab:simulation_studies_2_Type_I_error_rate}
\end{table}

Similar to the cases considered in Section \ref{sec:sim_correct_model}, Table \ref{tab:simulation_studies_2_g_computation_inferences} demonstrates that the g-computation point estimators under the Omitted Covariates and Random Errors scenarios exhibit negligible bias for estimating $\Delta_{\mathrm{OR}}, \Delta_{\mathrm{RR}}$, and $\Delta_{\mathrm{RD}}$. In addition, the nonparametric bootstrap confidence intervals from PROCOVA-LR have smaller widths, on average, compared to those from the unadjusted models, with the ratio of the confidence intervals' widths approximately equal to the expected efficiency factor. Thus, for these two scenarios with a misspecified PROCOVA-LR model, the efficiency factor corresponds to the expected gain from PROCOVA-LR in terms of inferential precision.

\begin{table}[h]
    \centering
    \footnotesize
    \begin{tabular}{|c | c | c | c | c | c | c | c | c|}
        \hline
                 & Estimand & Estimand & \multicolumn{2}{c|}{Deviations of} & \multicolumn{3}{c|}{Width of} & $f_{\mathrm{EFF}}$ \\
        &  & Value & \multicolumn{2}{c|}{Estimators} & \multicolumn{3}{c|}{$95\%$ CI} & \\
        \hline 
        Scenario & & & UN & P-LR & UN & P-LR & Ratio & \\
        \hline
        \hline
        Omitted & $\Delta_{\mathrm{OR}}$ & $1.48$ & $-.019$ & $-.007$ & $.85$ & $.69$ & $.81$ & $.84^{\dag}$ \\
        Covariate & & $(.01)$ & $(.222)$ & $(.180)$ & $(.13)$ & $(.09)$ & $(.07)$ & $(.01)$ \\
         & $\Delta_{\mathrm{RR}}$ & $1.19$ & $-.004$ & $-.001$ & $.31$ & $.25$ & $.82$ & $.84^{\dag}$\\
         & & $(.01)$ & $(.081)$ & $(.066)$ & $(.03)$ & $(.02)$ & $(.04)$ & $(.01)$ \\
         & $\Delta_{\mathrm{RD}}$ & $.096$ & $-.0003$ & $.0007$ & $.137$ & $.112$ & $.82$ & $.84^{\dag}$ \\
         & & $(.002)$ & $(.0361)$ & $(.0298)$ & $(.002)$ & $(.003)$ & $(.02)$ & $(.01)$ \\
         \hline
        Random & $\Delta_{\mathrm{OR}}$ & $1.73$ & $-.038$ & $-.030$ & $1.49$ & $1.33$ & $.90$ & $.90^{\dag\dag}$ \\
        Error & & $(.02)$ & $(.364)$ & $(.332)$ & $(.34)$ & $(.28)$ & $(.09)$ & $(0.01)$ \\
        & $\Delta_{\mathrm{RR}}$ & $1.16$ & $-.002$ & $-.001$ & $.26$ & $.23$ & $.90$ & $.90^{\dag\dag}$ \\
        & & $(.01)$ & $(.065)$ & $(.060)$ & $(.02)$ & $(.02)$ & $(.04)$ & $(0.01)$ \\
        & $\Delta_{\mathrm{RD}}$ & $.108$ & $.0002$ & $.0004$ & $.155$ & $.140$ & $.91$ & $.90^{\dag\dag}$ \\
        & & $(.003)$ & $(.0394)$ & $(.0364)$ & $(.004)$ & $(.004)$ & $(.02)$ & $(0.01)$ \\
        \hline
    \end{tabular}
    \caption{Summary of the frequentist properties for the g-computation based point estimators and confidence intervals for the marginal estimands under the Omitted Covariate and Random Error scenarios. The standard deviations of the mean values are indicated in parentheses below the means. The small values for the average deviances indicate that the g-computation based point estimators exhibit negligible bias for estimating $\Delta_{\mathrm{OR}}, \Delta_{\mathrm{RR}}$, and $\Delta_{\mathrm{RD}}$. The widths of the confidence intervals from PROCOVA-LR are shorter than those from the unadjusted analysis, and the efficiency gains are closely aligned with the efficiency factor values. $\dag$ The efficiency factor for this case is calculated using $m_i$ only, omitting $x_i$. $\dag\dag $ The expected efficiency factors are calculated based on the adjusted efficiency factor in equation (\ref{eq:factor_adj}).}
    \label{tab:simulation_studies_2_g_computation_inferences}
\end{table}

\normalsize

\section{Concluding Remarks}
\label{sec:disc}

The design and analysis of RCTs with binary endpoints has traditionally been a complicated endeavor due to non-collapsibility. Our PROCOVA-LR methodology helps to resolve this challenge. PROCOVA-LR incorporates a covariate adjustment in the logistic regression analysis by means of an AI-algorithm that is pre-trained on historical control data. It controls the Type I error of the Wald test on the conditional odds ratio estimand, and yields consistent estimators for the marginal risk difference, relative risk, and odds ratio estimands via g-computation. We derived prospective formulae that enable one to quantify the expected benefits, in terms of power gain and/or sample size reduction, of PROCOVA-LR for testing the conditional odds ratio. The scope of our formulae was extended to g-computation based inferences for marginal estimands under PROCOVA-LR. These formulae are a function of the average and variance of the probabilities of an event under control that incorporate the prognostic scores across the RCT participants. They remain applicable even in cases of model misspecifications, as demonstrated via our simulation studies. This corresponds to the fact that g-computation is robust against model misspecifications, more scientifically justifiable (as it considers potential outcomes), and more flexible in terms of enabling one to consider the multiple types of estimands $\Delta_{\mathrm{RD}}, \Delta_{\mathrm{RR}}$, and $\Delta_{\mathrm{OR}}$. These marginal estimands are generally acceptable by regulators, although the choice of estimand ultimately requires discussion with regulators.

It is important to recognize the practical distinction between conditional and marginal estimands. Patients and physicians arguably care more about individualized treatment effects in medical practice, which can only be obtained via a covariate adjusted model. However, decision-making for the totality of a patient population (e.g., a benefit-risk assessment of a new treatment) requires consideration of marginal treatment effects. \citet{schulz_2010} describes how health authorities are advised to report both marginal and conditional estimands. Furthermore, inferences for the conditional estimand from logistic regression do not explicitly provide information for each treatment level, as they only provide estimates for comparison. Such estimates may not provide sufficient clinical information when there is no explicit reference information. We believe that the best solution to meet both the individualized and population-level perspectives is to specify the PROCOVA-LR model so as to enable individual-level estimators, and the integration of the individualized treatment effects to obtain population-level estimators. PROCOVA-LR thus enables interpretable inferences on both conditional and marginal estimands, and can enable the design of smaller and faster RCTs whose primary endpoints are binary, by leveraging the power of modern AI for covariate adjustment.

\clearpage
\bibliographystyle{apalike}
\bibliography{references}

\clearpage

\appendix

\section{Correspondence Between Bias Factors, Asymptotic Relative Efficiencies, and Efficiency Factors}
\label{app:additional_simulation}

\begin{table}[h]
    \centering
    \begin{tabular}{|c | c | c | c | c | c|}
        \hline
        Scenario & $\widehat{\beta_1^*}/\hat{\beta}_1$ & $\mathrm{Var}(\widehat{\beta_1^*})/\mathrm{Var}(\hat{\beta}_1)$ & $E(\mu_{0,i})$ & $\mathrm{Var}(\mu_{0,i})$ & $f_{\mathrm{EFF}}$ \\
        \hline \hline
        Baseline & $0.71$ & $0.71$ & $0.67$ & $0.06$ & $0.85$ \\
           & $(9.88)$ & $(0.04)$ & $(0.01)$ & -  & $(0.01)$ \\
        \hline
        Large & $0.73$ & $0.71$ & $0.67$ & $0.06$  & $0.85$ \\
        Effect & $(0.23)$ & $(0.04)$ & $(0.01)$ & - & $(0.01)$ \\
        \hline
        Large & $0.54$ & $0.52$ & $0.63$ & $0.11$ & $0.73$  \\
        Variance & $(1.56)$ & $(0.04)$ & $(0.01)$ & -  & $(0.01)$ \\
        \hline
        High & $0.68$ & $0.67$ & $0.83$ & $0.05$ & $0.82$ \\
        Prevalence & $(0.41)$ & $(0.04)$ & $(0.01)$ & -  & $(0.01)$ \\
        \hline
        Omitted & $0.67$ & $0.66$ & $0.50$ & $0.07$ & $0.84^{\dag}$ \\
        Covariate & $(0.52)$ & $(0.03)$ & $(0.01)$ & - & $(0.01)$ \\
        \hline
        Random & $0.83$ & $0.81$ & $0.67$ & $0.06$  & $0.90^{\dag\dag}$ \\
        Error & $(3.05)$ & $(0.04)$ & $(0.01)$ & -  & $(0.01)$ \\
        \hline
        Shift and & $0.82$ & $0.81$ & $0.67$ & $0.06$ & $0.90^{\dag\dag}$ \\
        Random Error & $(0.47)$ & $(0.04)$ & $(0.01)$ & - & $(0.01)$ \\
        \hline
    \end{tabular}
    \caption{The bias factors, AREs, averages of the probability of an event under control, variances of the probability of an event under control, and the efficiency factors for the simulation scenarios in Section \ref{sec:sim}. $\dag$ The efficiency factor for this case is calculated using $m_i$ only, omitting $x_i$. $\dag\dag $ The expected efficiency factors are calculated based on the adjusted efficiency factor in equation (\ref{eq:factor_adj}).}
    \label{tab:app_bias_factor}
\end{table}

\end{document}